\documentstyle[preprint,prd,aps]{revtex}
\begin{document}

\vspace{1truein}

\preprint{
\vbox{\halign{&##\hfil\cr
		& OHSTPY-HEP-T-97-001	\cr
		& January 1997	\cr
                &\vspace{0.6truein} \cr
}}}

\vspace{1truein}

\title{ Calculation of P-Wave Charmonium Decay Rates\\
 Using  Dimensional Regularization }

\author{Eric Braaten and Yu-Qi Chen}
\address{Physics Department, Ohio State University, Columbus OH 43210}

\maketitle

\begin{abstract}
Contributions to the decay rates of P-wave charmonium states that 
are proportional to $n_f \alpha_s^3$, where $n_f$ is the number of 
flavors of light quarks, are calculated  in the framework of 
nonrelativistic QCD using the threshold expansion method. 
Dimensional regularization is used to regularize the infrared
divergences that arise from the emission of a soft gluon. 
Our results are consistent with the original calculations of Barbieri et al.
\end{abstract}


\vfill \eject

\narrowtext

\section{Introduction}
The nonrelativistic QCD (NRQCD) factorization formalism\cite{B-B-L} provides 
a systematic framework for analyzing annihilation decay rates and inclusive 
production rates of heavy quarkonium. These rates are factored into 
short-distance coefficients and  long-distance NRQCD matrix elements. The 
short-distance coefficients can be calculated using  perturbative QCD as a 
power series in the strong coupling constant $\alpha_s(m_c)$ at the scale 
of the heavy quark mass. 
The matrix elements scale in a definite way with $v$, the typical relative 
velocity of the heavy quark.  The NRQCD factorization formalism therefore  
organizes the decay rate or production rate into a double expansion 
in powers of  $\alpha_s$  and $v$.

The {\it threshold expansion method}\cite{Braaten-Chen} 
is a general  method for calculating the short-distance coefficients
which fully exploits the NRQCD factorization formalism.
In this method, a quantity that is closely related to the 
creation or annihilation rate
of a $c\bar{c}$ pair near threshold 
is calculated using perturbation theory in full QCD and then 
expanded in powers of the relative 3-momentum ${\bf q}$ of the 
$c$ and $\bar{c}$. Matrix elements of 4-fermion operators in  NRQCD
are also calculated using perturbation theory and expanded 
around the threshold ${\bf q}= 0$. 
The short-distance coefficients in the factorization formula
are then determined by matching  these expansions
in ${\bf q }$ order by order in $\alpha_s$.
Finally, the NRQCD matrix elements for specific quarkonium states are
simplified by using rotational symmetry, heavy-quark 
spin symmetry, and the vacuum-saturation approximation.

In calculating the short-distance coefficients beyond 
leading order in $\alpha_s$, ultraviolet divergences and infrared 
divergences inevitably arise and need to be regularized.  In perturbative 
calculations, the most convenient method for regularizing 
both ultraviolet and infrared divergences is dimensional regularization.
We  recently generalized the 
threshold expansion  method to $N$ spatial dimensions,
so that dimensional regularization can be used consistently
in quarkonium calculations \cite{Braaten-Chen:dimreg}.
The QCD side of the matching condition and the matrix elements on the 
NRQCD side are calculated in $N$ dimensions using perturbation theory
and then matched to obtain 
the short-distance coefficients.  After renormalization of 
coupling constants in QCD and NRQCD, the short-distance coefficients 
may have poles in $N-3$, which must be removed
by renormalization of the 4-fermion operators in NRQCD.
One must take care to avoid simplifying the matrix elements of these 
operators using identities that are specific to 3 dimensions until
after these renormalizations have been carried out.
We illustrated our method by calculating gluon fragmentation
functions to S-wave and P-wave  charmonium states, resolving a 
discrepancy between two previous calculations of the fragmentation
function for $g \to \chi_{cJ}$.   

There are also discrepancies in the literature between various calculations 
of the annihilation decay rates for P-wave states.  In the original 
calculations by Barbieri et al. \cite{Barbieri:chi02,Barbieri:chi1}, 
either the binding energy of the $c\bar{c}$ pair or the energy 
resolution of a gluon jet was used as the infrared cutoff. 
In recent calculations by Huang and Chao \cite{Huang-Chao}
and by Petrelli \cite{Petrelli:PhD}, 
dimensional regularization was used for the infrared cutoff. 
The discrepancies between their results and those of Barbieri et al.
appear in the 
terms proportional to $n_f \alpha_s^3$, where $n_f$ is the number 
of light quark flavors.  This is the term that is sensitive to the infrared 
cutoff. In this paper, we  use the threshold expansion method 
in conjunction with dimensional regularization to calculate 
the term  proportional to $n_f \alpha_s^3$.  We find that
the discrepancy between the results of Barbieri et al. and the recent 
calculations using dimensional regularization is due simply to different
definitions of a color-octet NRQCD matrix element.  

The outline of the rest of this paper is as follows. In Section II,
we review the threshold expansion method 
in $N$ dimensions as it applies to decay rates. 
In Section III,
we  calculate the terms proportional to  $n_f \alpha_s^2$ in the 
short-distance coefficients of color-octet matrix elements and 
the terms proportional to  $\alpha_s^2$ and $n_f \alpha_s^3$ in the 
short-distance coefficients of color-singlet matrix elements.
In Section IV, we apply our results to the decay rates 
of spin-singlet S-wave and spin-triplet P-wave states.
Finally, we compare our results for the P-wave states with 
the previous calculations in the literature.

\section{Threshold Expansion Method}

The threshold expansion method for calculating the short-distance 
coefficients in inclusive production cross-sections was developed in 
Ref. \cite{Braaten-Chen} and generalized to $N$ spatial dimensions in 
Ref.~\cite{Braaten-Chen:dimreg}.
In this section, we review this method as it applies to annihilation decay
rates, since many of the formulas differ a little from the production 
case. 

The annihilation decay rate for  the charmonium state  $H$ 
can be written in the  factorized form \cite{B-B-L}
\begin{equation}
\Gamma ( H )
\;=\; 
{1 \over 2 M_{H} } \; 
\sum_{mn} {C_{mn}(\mu) \over m_c^{d_{mn}-N-1}} 
     	\langle H | {\cal O}_{mn} | H \rangle^{(\mu)} \;,
\label{fact-Gam}
\end{equation}
where $M_H$ is the  mass of the  state $H$ and $d_{mn}$ is the 
mass dimension of the operator ${\cal O}_{mn}$. 
The matrix elements $ \langle H | {\cal O}_{mn} | H \rangle $ 
are expectation values in the quarkonium state $H$ of local 4-fermion 
operators that have the structure
\begin{equation}
{\cal O}_{mn}
\;=\; \psi^\dagger {{\cal K}'}_m^\dagger \chi \;
	\chi^\dagger {\cal K}_n \psi \;,	
\label{O_mn}
\end{equation}
where $\psi$ and $\chi$ are the field operators for the heavy quark and 
antiquark in NRQCD, and ${\cal K}_n$ and ${{\cal K}'}_m^\dagger$
are products of a color matrix ($1$ or $T^a$), 
a spin matrix, and a polynomial in the gauge covariant 
derivative ${\bf D}$ in $N$ dimensions. The spin matrix is either the 
unit matrix or a polynomial in the Pauli matrices $\sigma^i$. The Pauli 
matrices in $N$ dimensions satisfy the anticommutation relations
$\{ \sigma_i, \sigma_j \} =  2 \delta_{ij}$, $i,j=1,\ldots, N$.
In 3 dimensions, they also satisfy the commutation relations
$[ \sigma_i, \sigma_j ] =  2 i\, \epsilon_{ijk} \sigma_k$, $i,j,k=1,2,3$.
The commutation relations can be used together with the  anticommutation 
relations to reduce all spin matrices to a linear 
combination of $1$ and $\sigma^i$.
However, since the commutation relations  are specific to $N=3$ dimensions, 
they should be used to simplify 
NRQCD matrix elements only after all poles in $N-3$
have been removed from the short-distance coefficients.

For the purpose of calculating short-distance coefficients,
it is convenient to define the states 
$ |H \rangle = | H ( {\rm\bf P } = 0 ) \rangle $
in (\ref{fact-Gam}) so that they have the 
standard relativistic normalization: 
\begin{equation}
\Big \langle H({\bf P}') \Big | H({\bf P}) \Big\rangle 
\;=\; 2 E_P \; (2 \pi)^N  \delta^N ({\bf P} - {\bf P}') \;,	
\label{norm-H}
\end{equation}
where $E_P = \sqrt{M_H^2 + {\bf P}^2 }$. 
With this choice of normalization,
the matrix elements in (\ref{fact-Gam}) differ from the
standard NRQCD matrix elements 
$\langle H | {\cal O}_1(^{2S+1}L_J)| H \rangle$ 
and $\langle H | {\cal O}_8(^{2S+1}L_J)| H \rangle$ introduced in 
Ref. \cite{B-B-L}.  The relation between them is discussed in 
Appendix B of \cite{Braaten-Chen}.  
Up to corrections of relative order $v^2$,
the difference is a simple multiplicative factor.
The renormalization of the operators ${\cal O}_{mn}$ 
makes them depend on the renormalization 
scale $\mu$ of NRQCD, as indicated by the superscript $(\mu)$
on the matrix elements in (\ref{fact-Gam}).  
This dependence will be suppressed whenever it is not essential.

The factors of $m_c$ in (\ref{fact-Gam}) are chosen so that the 
short-distance coefficients $C_{mn}$ are dimensionless.  
The dimension of the operator ${\cal O}_{mn}$ in (\ref{O_mn}) is
$d_{mn} = 2N+D$, where $D$ is the number of covariant derivatives 
in the operator.
Since the coefficients $C_{mn}$ take into account the effects of 
short distances of order $1/m_c$, they
can be calculated as perturbation series in the QCD coupling 
constant $\alpha_s(2m_c)$.  The coefficients in the perturbation
series depend on $\ln(\mu/m_c)$ in such a way as to cancel
the $\mu$-dependence of the matrix elements. 

The short-distance coefficients for annihilation decay rates
can be determined by matching perturbative calculations of 
$c \bar{c}$ scattering amplitudes. 
Let $c \bar c({\bf q},\xi,\eta)$ represent a state that consists of a $c$ 
and a $\bar c$ with spatial momenta $\pm {\bf q}$ in the 
$c \bar c$ rest frame and spin and color states specified by the spinors
$\xi$ and $\eta$. The standard relativistic normalization is
\begin{equation}
\Big \langle c({\bf q}_1',\xi') \bar c({\bf q}_2',\eta')
	 \Big | c({\bf q_1},\xi) \bar c({\bf q_2},\eta)  \Big \rangle 
\;=\; 4 E_{q_1} E_{q_2} \; (2 \pi)^{2 N}  \delta^N ({\bf q}_1 - {\bf q}_1')
	\delta^N({\bf q}_2 - {\bf q}_2') \; 
	\xi'^\dagger \xi {\eta'}^\dagger \eta \; ,
\label{cc-norm}
\end{equation}
where $E_q = \sqrt{m_c^2 + {\bf q}^2}$.  The spinors are normalized so that 
 $\xi^\dagger \xi = 1$, and similarly for $\eta$, 
$\xi'$, and $\eta'$.
 Using the abbreviated notation 
$c \bar c \equiv c \bar c({\bf q},\xi,\eta)$ and
$c \bar c' \equiv c \bar c({\bf q}',\xi',\eta')$, the matching condition
 in the threshold expansion method of Ref. \cite{Braaten-Chen} is 
\begin{eqnarray}
&& \sum_X (2 \pi)^{N+1} \delta^{N+1}( P - k_X) \;
	({\cal T}_{ c \bar c' \to  X})^* \; 
	{\cal T}_{ c \bar c \to  X} \Big|_{pQCD}
\nonumber \\ 
&& \hspace{1.5in} 
\;=\; \sum_{mn} {C_{mn}(\mu) \over m_c^{d_{mn}-N-1}} \;
	\langle c \bar{c}' | \psi^\dagger {{\cal K}'}_m^\dagger \chi \; 
	\chi^\dagger {\cal K}_n \psi | c \bar{c} \rangle^{(\mu)} 
	\Big|_{pNRQCD} \;,
\label{match}
\end{eqnarray}
where $P=(2 E_q, 0 ) $ is the momentum of the $c\bar{c}$ 
pair, ${\cal T}_{ c \bar c \to  X}$ is the $T$-matrix element for annihilation 
of the  $c \bar{c}$ into the state $X$ consisting of light partons,
  and $k_X$ is the sum of the momenta 
of the outgoing partons.
The sum  over $X$ on the left side of (\ref{match}) includes integration
over the phase space of the final state partons and sum over their spin and 
color quantum numbers. By the optical theorem,
the left side of  (\ref{match}) is proportional to   the imaginary 
part of the  annihilation contribution to the $T$-matrix element 
${\cal T}_{ c \bar c \to c \bar c' }$.  
Specifically, it is the sum of all cut diagrams for 
which the cut does not pass through any heavy quark lines. 

To carry out the matching procedure, the left side of (\ref{match}) is
calculated using perturbation theory in full QCD, and then expanded 
in powers of ${\bf q}$ and ${\bf q}'$. The matrix elements on the right 
side of (\ref{match}) are 
calculated using perturbation theory in NRQCD, and then expanded 
in powers of ${\bf q}$ and ${\bf q}'$.  The short-distance coefficients
$C_{mn}$ are obtained by matching the terms in the expansions
in ${\bf q}$ and ${\bf q}'$ order by order in $\alpha_s$. 

The calculation of the short-distance coefficients 
can be simplified by averaging 
both sides of the matching condition (\ref{match}) over rotations of the 
vectors ${ \rm\bf q}'$ and ${ \rm\bf q}$  and the spinors 
$\xi$, $\eta$, $\xi'$, and $\eta'$ that specify the states  of the initial
and final 
$c \bar c$ pairs. On the NRQCD side of the matching equation, the average over 
rotations can be accomplished simply by restricting the operators 
${\cal O}_{mn}$ to be rotationally invariant. 

In the perturbative calculations of the matching condition 
(\ref{match}), infrared and ultraviolet divergences can appear 
on both sides of the equation.  Since NRQCD is constructed to be 
equivalent to full QCD at low momenta, the infrared divergences 
on both sides must match.  They therefore cancel in the 
short-distance coefficients $C_{mn}$.  Any ultraviolet divergences
on the left side are eliminated by renormalization 
of the QCD coupling constant and the heavy quark mass.
On the right side,  the ultraviolet divergences are eliminated
by renormalization of the gauge coupling constant and other parameters 
in the NRQCD lagrangian and 
by renormalization of the 4-fermion operators of NRQCD.
 
The matching calculations are particularly simple if dimensional 
regularization is used to regulate both infrared and ultraviolet divergences.
Radiative corrections to the NRQCD matrix elements 
vanish identically, because after expanding 
the integrand of the radiative correction 
in powers of ${\bf q}$ and ${\bf q}'$, there is no
momentum scale in the dimensionally regularized integral. 
The radiative corrections to the matrix elements do include infrared 
poles in $\epsilon= ( 3-N)/2$ 
that match the infrared divergence on the QCD side of the matching condition,  
but they are  canceled by  ultraviolet poles in $\epsilon$.
Thus the only noncanceling
contributions to the NRQCD side of the matching condition
are the tree-level contributions of the matrix elements, including those
matrix elements that arise from counterterms 
associated with operator renormalization.

\section{Perturbative matching of QCD and NRQCD}

In this section, we use the threshold expansion method to calculate
selected terms in the short-distance coefficients in the factorization 
formula (\ref{fact-Gam}).
We calculate the term proportional to $n_f \alpha_s^2$ from 
the color-octet annihilation processes $c\bar{c} \to q \bar{q}$ 
and the terms proportional to $\alpha_s^2$ and $n_f \alpha_s^3$ 
from the color-singlet annihilation processes 
$c\bar{c} \to q \bar{q} g $ and $ c \bar{c} \to g g$. 
Dimensional regularization is used as a cutoff for both infrared 
and ultraviolet divergences.

\subsection{ Color-octet terms from $c\bar{c} \to q \bar{q}$ }

The terms on the QCD side of the matching condition (\ref{match}) 
that are proportional to  $n_f\alpha_s^2$ come from 
the annihilation  process $c \bar{c} \to q \bar{q}$. 
For these terms, the left side of (\ref{match}) reduces to
\begin{equation}
\int_{\ell_1} \int_{\ell_2} ( 2\pi )^{N+1} \delta^{N+1}( P - \ell_1 - \ell_2 )
\sum ( {\cal T}_{c\bar{c}' \to q \bar{q} } )^*
       {\cal T}_{c\bar{c} \to q \bar{q} } \;,
\label{match-qqbar}
\end{equation}
where $\ell_1$ and $\ell_2$ are the momenta of the $q$ and $\bar{q}$ and
the remaining sum is over their colors, spins, and flavors.  
The symbol $\int_k$ denotes the 
$N$-dimensional integral over the phase space associated with the
momentum $k$:
\begin{equation}
\int_k  \;\equiv\; \int {d ^N k \over (2 \pi)^N 2 k_0} \;.
\end{equation}

The term in (\ref{match-qqbar}) that is proportional to $n_f \alpha_s^2$
is given by the cut Feynman diagram in Figure 1. 
The T-matrix element for 
$c(p) \bar{c}( \bar{p}) \to q (l_1) \bar{q}(l_2)$ is 
\begin{equation}
{\cal T}_{c\bar{c} \to q \bar{q} } = 
 \bigg( g_s \mu^{\epsilon} \bar{v}(\bar{p}) \gamma^\mu T^a u(p) \bigg)
 \;{1 \over 4 E_q^2 } \;
 \bigg( g_s \mu^{\epsilon} \bar{u}(l_1) \gamma_\mu T^a v(l_2) \bigg) \;.
\label{T-qqbar}
\end{equation}
The $c$ and $\bar{c}$ have 4-momenta $p=( E_q, {\rm\bf q} ) $ and  
$\bar{p} =( E_q, -{\rm\bf q} ) $ where $E_q = \sqrt{m_c^2 + {\bf q}^2 }$.  
The coupling constant in (\ref{T-qqbar}) has been written
$g_s \mu^\epsilon$, where $\epsilon = (3-N)/2$ and $\mu$ is the 
scale of dimensional regularization, so that $g_s$ remains dimensionless
in $N$ dimensions. 
The integrals over $\ell_1$ and $\ell_2$ in (\ref{match-qqbar})
can be carried out by using 
the  energy-momentum-conserving delta function: 
\begin{eqnarray}
&& \int_{\ell_1} \int_{\ell_2}  \; 
   (2\pi)^{N+1} \delta^{N+1} ( P - l_1 -l_2 )
  \sum 
  \bar{v}(l_2) \gamma_\nu T^b u(l_1) \bar{u}( l_1 ) \gamma_\mu T^a v( l_2 )  
\nonumber \\
&& \hspace{1.5in} \;= \;
  n_f
{(N-1)\, \Gamma({3 \over 2 } )  
	\over 8  \pi N \Gamma( {N  \over 2 })} \; 
  \left( {P^2 \over 16 \pi } \right ) ^ { - \epsilon} \;
  \left( -P^2 g_{\mu \nu} + P_\mu P_\nu \right) \delta^{ab} \;,
\label{octet-QCD}
\end{eqnarray}
where $ P = (2 E_q, {\bf 0})$. We have expressed this in Lorentz-invariant 
form for later use.
Inserting this into the QCD side of the matching condition, it reduces to
\begin{equation}
 - n_f 
{(N-1)\, \Gamma({3 \over 2 } )  
	\over 32 \pi N \Gamma\left( {N  \over 2 } \right) } \; 
  g_s^4 \mu^{  4  \epsilon } \;
\left( {E_q^2 \over 4 \pi } \right ) ^ { - \epsilon}  {1 \over E_q^2}
\bar{u}(p) \gamma^\mu T^a v(\bar{p}) \bar{v}(\bar{p}) \gamma_\mu T^a u(p) \;. 
\label{qqbar:1}
\end{equation}

Using the formulas in Appendix A, $\bar{v}(\bar{p}) \gamma_\mu T^a u(p)$
can be  expressed 
in terms of nonrelativistic Pauli spinors. Expanding to linear order in 
${\rm\bf q}$, we obtain 
\begin{equation}
\bar{v}(\bar{p}) \gamma ^\mu T^a u (p) 
\;\approx \; - 2 m_c   \; g^{\mu \,i} \; \eta^\dagger \sigma^i T^a \xi \;.
\end{equation}
The QCD side of the matching condition (\ref{qqbar:1}) then  reduces to  
\begin{equation}
n_f \, 
{2 \pi \, (N-1)\, \Gamma({3 \over 2 })  
	\over N \Gamma({N \over 2}) } \; 
  \alpha_s^2 \mu^{  4  \epsilon } \;
  \left( {m_c^2 \over 4 \pi } \right ) ^ { - \epsilon} \;
\xi'^\dagger \mbox{\boldmath $\sigma$} T^a \eta' 
	\cdot \eta^\dagger \mbox{\boldmath $\sigma$} T^a \xi \;.
\label{qqbar:2}
\end{equation}
Setting $N=3$, this simplifies to
\begin{equation}
{4 \pi n_f \over 3 } \; \alpha_s^2 \;
\xi'^\dagger \mbox{\boldmath $\sigma$} T^a \eta' 
	\cdot \eta^\dagger \mbox{\boldmath $\sigma$} T^a \xi \; .
\label{qqbar:3}
\end{equation}
Dividing by $(2 m_c)^2$ to account for  the difference 
in the normalization of states, we reproduce twice the leading term in 
${\rm Im} {\cal M}$ given in (A13) of Ref. \cite{B-B-L}.

\subsection{ Color-singlet terms from $c \bar{c} \to q \bar{q} g $ }

Terms proportional to $n_f \alpha_s^3$  on the QCD side of the matching 
condition come from the processes $c \bar{c} \to q \bar{q} g$
and  $c \bar{c} \to q \bar q$. For the process $c \bar{c} \to q \bar{q} g$,
the left side of (\ref{match}) can be written
\begin{equation}
\int_{k} \int_{\ell_1} \int_{\ell_2} \;
( 2\pi )^{N+1} \delta^{N+1} ( P - k - \ell_1 - \ell_2 ) \;
\sum ( {\cal T}_{c\bar{c}' \to q \bar{q} g } )^*
       {\cal T}_{c\bar{c}  \to q \bar{q} g } \;,
\label{match-gqqbar}
\end{equation}
where $\ell_1$, $\ell_2$, and $k$ are the momenta of the 
$q$, $\bar q$, and gluon, respectively, and the remaining sum is over the 
colors, spins, and flavors of these partons. 
The term proportional to $n_f \alpha_s^3$ comes from the cut 
Feynman diagrams in Figure 2.
The T-matrix element in Feynman gauge for $c(p) \bar{c}( \bar{p} ) \to
q(l_1) \bar{q}(l_2) g (k) $ can be written
\begin{equation}
{\cal T}_{c \bar{c}  \to q \bar{q} g }
=
{ \cal A } ^{\mu a}_{ c \bar{c}  \to g^* g } \;
{ 1 \over ( \ell_1 + \ell_2 )^2 } \;
\bigg( g_s \mu^{\epsilon} \bar{u}(l_1) \gamma _{\mu} T^a v (l_2) \bigg)\;,
\label{T-gqqbar}
\end{equation}
where 
$ {\cal A}^{\mu a}_{c \bar{c}  \to g^* g } $ is 
the amplitude for the process 
$ c(p)\; \bar c(\bar{p}) \to g^*(l) g(k) $:
\begin{eqnarray}
{\cal A}^{\mu a}_{ c \bar c \to g^* g}
&\;= \;&
 {g_s^2 \mu^{2 \epsilon} } \; \epsilon_{\nu}^{b}(k) \; 
\bar v(\bar p) 
\Bigg[ \; {\gamma^\nu (\not \! k \; -  \not \! \bar p \;+ m_c) \gamma^\mu
	\over 2 \bar{p} \cdot k} \, T^b T^a 
\nonumber \\
&& \hspace{2in} 
\;+\; {\gamma^\mu (\not \! p \; -  \not \! k \;  + m_c) \gamma^\nu 
	\over 2 p \cdot k} \, T^a T^b \; \Bigg] u(p) \;.
\label{A-gqqbar} 
\end{eqnarray}
We will compute the matching condition for the case of color-singlet 
$c \bar{c}$ pairs.  The color
matrices $T^a T^b$ and $T^b T^a$ in (\ref{A-gqqbar}) 
can then be replaced by $\delta^{ab}/6$.
We will further simplify the matching condition by averaging 
the initial state  $c \bar c({\bf q},\xi,\eta)$ and the final state
$c \bar c({\bf q}',\xi',\eta')$ over the rotation group.
 
The integral over $\ell_1$ and $\ell_2$ in (\ref{match-gqqbar})
can be carried out by inserting the identity
\begin{equation}
\int { d \ell^2 \over 2 \pi } 
\int_\ell \; (2 \pi)^{N+1} \delta^{N + 1} (\ell - \ell_1 - \ell_2) 
\;=\; 1\;.
\end{equation}
After using (\ref{octet-QCD}) with $P$ replaced by $\ell$ to integrate over 
$\ell_1$ and $\ell_2$, we obtain
\begin{eqnarray}
&&
n_f \; 
{(N-1) \Gamma({3 \over 2}) \over 8  \pi N \Gamma({N \over 2})} \;
g_s^2 \mu^{2 \epsilon} \int {d \ell^2 \over 2 \pi} \int_{\ell} \int_k \;
(2 \pi)^{N+1} \delta^{N+1}(P - \ell - k) \;
\nonumber \\
&& \hspace{1in}\times  
{1 \over (\ell^2)^2}
	\left( {\ell^2 \over 16 \pi } \right ) ^ {- \epsilon} 
	\left( - \ell^2 g_{\mu \nu} + \ell_\mu \ell_\nu \right)   
\sum \left( {\cal A}^{\nu a}_{ c\bar{c}' \to g^* g} \right)^* 
	{\cal A}^{\mu a}_{ c\bar{c}  \to g^* g} \;.
\label{gqqbar:1}
\end{eqnarray}
The remaining sum is over the colors and spins of the real gluon. 
 The $\ell_\mu \ell_\nu $ term gives no
contribution due to current conservation. 

We proceed to carry out the nonrelativistic expansion of the factor 
in (\ref{gqqbar:1}) that 
involves the amplitude ${\cal A}^{\mu a}_{ c\bar{c}  \to g^* g}$.
Using the identities in Appendix A, we expand (\ref{A-gqqbar}) 
to linear order in ${\bf q}$: 
\begin{eqnarray}
{\cal A}^{\mu a}_{ c \bar c \to g^* g} 
&\;=\;& {g_s^2 \mu^{2 \epsilon} \over 12 m_c k_0 } \; \epsilon_{\nu}^{ a}(k)
\Bigg\{   \; m_c \; g^{\mu i}  g^{\nu j} \;
	\eta^\dagger \{ [ \sigma^i, \sigma^j ], 
	{\bf k} \cdot \mbox{\boldmath $\sigma$} \} \xi
\nonumber \\
&&  \hspace{1in} 
\;+\; { 4 } \Bigg[ 
 {1 \over  k_0  } \, { k}^i 
	\left( k^\mu g^{\nu j} + g^{\mu j} k^\nu -2m_c g^{\mu j} g^{\nu 0} 
		- { k}^j \, g^{\mu \nu} \right) 
\nonumber \\ 
&& \hspace{1.5in} 
\;-\; { k}^i ( g^{ \mu 0 }  g^{\nu j} 
             - g^{ \nu 0 }  g^{\mu j} ) 
\;+\; { k}^j ( g^{ \mu 0 }  g^{\nu i} 
             - g^{ \nu 0 }  g^{\mu i} ) 
\nonumber \\ 
&& \hspace{1.5in} 
\;+\;  k_0    \; g^{\mu i}  g^{\nu j} 
\;+\;  (2 m_c - k_0) \; g^{\mu j}  g^{\nu i} 
\; \Bigg] \; \eta^\dagger q^i \sigma^j \xi \; \Bigg\} \; .
\label{A-gqqbar:NR} 
\end{eqnarray} 
In $N>3$ dimensions, the spin matrix 
$\{[\sigma^i,\sigma^j],\sigma^k\}$, 
which is totally antisymmetric in its three indices, is linearly 
independent of 1 and $\sigma^i, i=1,\ldots,N$.  In 3 dimensions, 
it reduces to the unit matrix multiplied by $-4i \epsilon^{ijk}$. 

After multiplying ${\cal A}^{\mu a}_{ c\bar{c}  \to g^* g}$
by $({\cal A}^{\nu a}_{ c\bar{c}  \to g^* g})^*$,
the spinor factors can be 
simplified by averaging the vectors ${\bf q}$ and  ${\bf q}'$
and the spinors $\xi$, $\eta$, $\xi'$, and $\eta'$ over the rotation group.  
The factor 
$\xi'^\dagger q'^k \sigma^l \eta' \eta^\dagger q^i \sigma^j \xi$ 
reduces to a linear combination of three independent spinor factors: 
\begin{eqnarray}
\overline{ 
\xi'^\dagger q'^k \sigma^l \eta' \eta^\dagger q^i \sigma^j \xi }
& = & {1 \over N(N-1)(N+2) } \; 
\nonumber \\
&& 
\times \, \Bigg[\; 
\left( (N+1) 
\delta^{ij} \delta^{kl} - \delta^{ik} \delta^{jl} - \delta^{il} \delta^{jk}  
\right ) 
\xi'^\dagger {\bf q}' \cdot \mbox{\boldmath $\sigma$} \eta' 
	\eta^\dagger {\bf q} \cdot \mbox{\boldmath $\sigma$} \xi 
\nonumber \\
&& \; + \; 
\left( (N+1) 
\delta^{ik} \delta^{jl} - \delta^{ij} \delta^{kl} - \delta^{il} \delta^{jk}  
\right ) 
\, \xi'^\dagger q'^m \sigma^n \eta' \eta^\dagger q^m \sigma^n \xi 
\nonumber \\
&&  \; + \; 
\left( (N+1) 
\delta^{il} \delta^{jk} - \delta^{ij} \delta^{kl} - \delta^{ik} \delta^{jl}  
\right ) 
\, \xi'^\dagger q'^m \sigma^n \eta' \eta^\dagger q^n \sigma^m \xi 
\;\Bigg] \;. 
\label{spinor-average}
\end{eqnarray}
The spinor factors 
$\xi'^\dagger q'^l \sigma^m \eta' 
	\eta^\dagger \{ [ \sigma^i, \sigma^j ], \sigma^k \} \xi$
and 
$\xi'^\dagger \{ [ \sigma^l, \sigma^m ], \sigma^n \} \eta' 
	\eta^\dagger q^i \sigma^j \xi$
average to zero, while 
$\xi'^\dagger \{ [ \sigma^l, \sigma^m ], \sigma^n \} \eta' 
	\eta^\dagger \{ [ \sigma^i, \sigma^j ], \sigma^k \} \xi$
reduces to a single spinor factor: 
\begin{equation}
\begin{array}{lcl}
& &
\overline{ 
\xi'^\dagger \{ [ \sigma^l, \sigma^m ], \sigma^n \} \eta' 
	\eta^\dagger \{ [ \sigma^i, \sigma^j ], \sigma^k \} \xi }
\\ 
&& \;= \;   \displaystyle
 {1 \over N(N-1)(N-2) } 
\;\; \left|\; 
\begin{array}{ccc}
  \delta^{il} ~ & ~ \delta^{jl} ~ & ~ \delta^{kl} \\ 
  \delta^{im} ~ & ~ \delta^{jm} ~ & ~ \delta^{km} \\ 
  \delta^{in} ~ & ~ \delta^{jn} ~ & ~ \delta^{kn} 
\end{array}
\; \right|\; \;
\xi'^\dagger \{ [ \sigma^r, \sigma^s ], \sigma^t \} \eta' 
  \eta^\dagger \{ [ \sigma^r, \sigma^s ], \sigma^t \} \xi \;.
\end{array}
\label{tensor-average}
\end{equation}
In 3 space dimensions, the spinor factor on 
the right side of (\ref{tensor-average}) can be simplified 
by using the commutation relations of the Pauli matrices:
\begin{equation}
\xi'^\dagger \{ [ \sigma^r, \sigma^s ], \sigma^t \} \eta' 
  \eta^\dagger \{ [ \sigma^r, \sigma^s ], \sigma^t \} \xi
\; = \;
-96 \; \xi'^\dagger \eta' \eta^\dagger \xi ,
\qquad N=3 \;.
\label{spinor-anti}
\end{equation}
However this simplification can only be made if there are no poles in 
$ N - 3 $ multiplying the spinor factor. 
After averaging over the rotation group
using (\ref{spinor-average}) and (\ref{tensor-average}) and
simplifying the Lorentz algebra, we obtain
\begin{eqnarray}
&& \left( - g_{\mu \nu} \right)   
\sum \overline{ \left( {\cal A}^{\nu a}_{ c\bar{c}' \to g^* g} \right)^* 
	{\cal A}^{\mu a}_{ c\bar{c}  \to g^* g} }
\;=\; {128 \pi^2 \over 9 N (N+2)} \; \alpha_s^2 \mu^{4 \epsilon}
	\; {1 \over x^2}
\nonumber \\
& & \hspace{0.5in}
\times \;\Bigg(\; 
6 (N+2) x^2 
\left( \mbox{$-{1 \over 96}$}
  \xi'^\dagger \{ [ \sigma^r, \sigma^s ], \sigma^t \} \eta' 
  \eta^\dagger \{ [ \sigma^r, \sigma^s ], \sigma^t \} \xi \right)
\nonumber \\
& & \hspace{1in}
\; + \; \left[ 4 (N+2) x + (N-1) x^2 \right]  \; {1 \over m_c^2} \;
    \xi'^\dagger {\bf q}'\cdot \mbox{\boldmath $\sigma$} \eta' 
    \eta^\dagger {\bf q} \cdot \mbox{\boldmath $\sigma$} \xi
\nonumber \\
& & \hspace{1in}
\; + \; \left[ 4 (N+2) (N-1) (1-x) + (2N^2+N-9) x^2 \right]  \; {1 \over m_c^2} \;
	\xi'^\dagger q'^m \sigma^n \eta' \eta^\dagger q^m \sigma^n \xi 
\nonumber \\
& & \hspace{1in}
\; + \; \left[ 4 (N+2)(N-2) x - (2N^2-N-7) x^2 \right]  \; {1 \over m_c^2} \;
	\xi'^\dagger q'^m \sigma^n \eta' \eta^\dagger q^n \sigma^m \xi 
\;\Bigg)\; ,
\label{AstarA}
\end{eqnarray}
where $x = k_0/m_c$.

At this point, the integrand of (\ref{gqqbar:1}) 
has been reduced to  a function of $\ell^2 = 4 E_q (E_q - k_0)$. 
We can therefore 
carry out the phase space integral over $ k $ and $ \ell $:
\begin{eqnarray}
&&
\int_{\ell} \int_k \;
( 2 \pi )^{ N+1} \delta ^{ N+1} ( P - \ell - k ) 
=\;
{ \Gamma({3 \over 2}) \over 8 \pi \Gamma({N \over 2}) }\; 
\left( {P^2 \over 16 \pi} \right)^{N-3 \over 2} \;  
\left( {P^2 - \ell ^2  \over P^2} \right) ^{N-2} \; .
\label{int-l-k}
\end{eqnarray}
Changing the integration variable to $x = 1 - \ell^2/(4 E_q^2)$
and taking the nonrelativistic limit, 
the QCD side of the matching condition reduces to 
\begin{eqnarray}
&&
n_f \; {(N-1) \Gamma^2({3 \over 2})
	\over 32 \pi^2 N \Gamma^2({N \over 2}) } \;
\alpha_s  \mu^{2 \epsilon} \;
\left({m_c^2 \over 4 \pi}\right)^{- 2 \epsilon} \;
\int^1_0 d x  \, x^{1 - 2 \epsilon }
     \left( {1-x} \right)^{-1 - \epsilon} 
\nonumber  \\ && \hspace{3in}
\times  ( - g_{\mu \nu }) \sum 
     \left( {\cal A}^{\nu a}_{ c\bar{c}' \to g^* g } \right)^* 
	    {\cal A}^{\mu a}_{ c\bar{c}  \to g^* g } \;. 
\label{gqqbar:2}
\end{eqnarray}

Infrared divergences arise from the endpoints of the integral 
in (\ref{gqqbar:2}).  
The infrared divergences from $x \to 1$
arise when the $q$ and $\bar{q}$ are collinear with the
virtual gluon, which is almost on-shell. 
They are canceled by collinear infrared divergences 
from radiative correction to the process 
$ c  \bar{c} \to g g $, which will be calculated in subsection IIIC.
The infrared divergence from $x \to 0 $, which 
appears only in the coefficient of 
$\xi'^\dagger q'^m \sigma^n \eta' \eta^\dagger q^m \sigma^n \xi$, 
arises when  the emitted  real gluon is soft.
It must be matched  by an infrared divergence 
on the NRQCD side of the matching condition.
The analytic expressions for the integrals over $x$ is
\begin{equation}
\int^1_0 d x \; x^{ n-1-2 \epsilon } ( 1 - x) ^{ -1 - \epsilon} 
\;=\;
{\Gamma(n - 2 \epsilon) \Gamma(- \epsilon) \over \Gamma(n - 3 \epsilon)} 
 \;. 
\label{int-x}
\end{equation}
Integrating over $ x $ and keeping only the terms that survive in the 
limit $\epsilon \to 0$, we obtain
\begin{eqnarray}
&& n_f
{4 (N-1)\, \Gamma^2({3 \over 2})
	\over 9 N^2 (N+2)  \Gamma^2({N \over 2})} \;
\alpha_s^3 \mu^{6 \epsilon} \; 
\left( { m_c^2 \over  4 \pi} \right)^{-2 \epsilon} \;
\nonumber \\
& & \hspace{1in}
\times { 1 \over \epsilon_{IR} }
\Bigg[ -6 (5 + 3 \epsilon) 
 \left( \mbox{$-{1 \over 96}$}
  \xi'^\dagger \{ [ \sigma^r, \sigma^s ], \sigma^t \} \eta' 
  \eta^\dagger \{ [ \sigma^r, \sigma^s ], \sigma^t \} \xi \right)
\nonumber \\
& & \hspace{2in}
\; - \; 2 (11 - 4 \epsilon) \, {1 \over m_c^2} \,
\xi'^\dagger {\bf q}' \cdot \mbox{\boldmath $\sigma$} \eta' 
	\eta^\dagger {\bf q} \cdot \mbox{\boldmath $\sigma$} \xi 
\nonumber \\
& & \hspace{2in}
\; - \;  2 (16 - 21 \epsilon) \, {1 \over m_c^2} \,
	\xi'^\dagger q'^m \sigma^n \eta' \eta^\dagger q^m \sigma^n \xi 
\nonumber \\
& & \hspace{2in}
\; - \;  2 (6 - 17 \epsilon) \, {1 \over m_c^2} \,
	\xi'^\dagger q'^m \sigma^n \eta' \eta^\dagger q^n \sigma^m \xi 
\; \Bigg] \; .
\label{gqqbar:3}
\end{eqnarray}
The subscript $IR$ on the pole in $\epsilon$ is a reminder that it
has an infrared origin.

\subsection{ Color-singlet terms from $c\bar{c} \to g  g $ } 

The remaining terms on the QCD side of the matching condition
that are proportional to $n_f \alpha_s^3$   
come from the process $c \bar{c} \to g g$.
For this process, the QCD side of the matching condition (\ref{match}) 
reduces to
\begin{equation}
{1 \over 2}\int_{k} \int_{\ell} (2 \pi)^{N+1} \delta^{N+1}(P - k - \ell)
\sum ( {\cal T}_{c\bar{c}' \to g g } )^*
       {\cal T}_{c\bar{c}  \to g g } \;,
\label{match-gg}
\end{equation}
where $k$ and $\ell$ are the momenta of the gluons, the sum is over their
spins and colors, and the factor of ${1 \over 2}$ comes from Bose statistics.

We first consider the QCD side of the matching condition at leading order 
in $\alpha_s$.  At this order, it is given by the cut Feynman diagrams in 
Figure 3.  The $T$-matrix element for 
$c(p) \bar{c}(\bar{p}) \to g(k) g (l) $ is
\begin{equation}
       {\cal T}_{c\bar{c}  \to g g }  
       \;=\; {\cal A}^{\mu a }_{c\bar{c}  \to g g } \; \epsilon^a_\mu (\ell) \;,
\label{T-gg}
\end{equation}
where ${\cal A}^{\mu a }_{c\bar{c}  \to g g } $ is identical to 
(\ref{A-gqqbar}) except that $\ell^2 = 0$. 
The nonrelativistic expansion of the matching condition is similar to 
but simpler than that for $c \bar c \to q \bar q g$, which was carried
out in subsection  IIIB.  The final result is 
\begin{eqnarray}
&&
{8 \pi \Gamma({3 \over 2})
	\over 9 N (N+2) \Gamma({N \over 2})} \;
\alpha_s^2 \mu^{4 \epsilon}  \; 
\left( {m_c^2 \over 4 \pi} \right)^{- \epsilon} \;
\nonumber \\
&& \hspace{1in}
\times \Bigg[\; 6(N + 2 ) \,  
\left( \mbox{$-{1 \over 96}$}
	\xi'^\dagger \{ [ \sigma^r, \sigma^s ], \sigma^t \} \eta' 
	\eta^\dagger \{ [ \sigma^r, \sigma^s ], \sigma^t \} \xi \right)
\nonumber \\
&& \hspace{1.5in}
\; + \; ( 5 N + 7 ) \, {1 \over m_c^2} \,
    \xi'^\dagger {\bf q}'\cdot \mbox{\boldmath $\sigma$} \eta' 
    \eta^\dagger {\bf q} \cdot \mbox{\boldmath $\sigma$} \xi
\nonumber \\
&&\hspace{1.5in}
\; + \; (2 N^2 + N - 9 ) \, { 1 \over m_c^2} \,
    \xi'^\dagger q'^m \sigma^n \eta' \eta^\dagger 
     ( q^m \sigma^n + q^n \sigma^m ) \xi 
\;\Bigg] \;.
\label{ggLO:1}
\end{eqnarray}
Setting $N=3$, this simplifies to 
\begin{eqnarray}
&& 
{16 \pi \over 135} \; \alpha_s^2\; 
\bigg[\;
15 \, \xi'^\dagger \eta' \eta^\dagger \xi 
\; + \; 11 \, {1 \over m_c^2} \,
    \xi'^\dagger {\bf q}'\cdot \mbox{\boldmath $\sigma$} \eta' 
    \eta^\dagger {\bf q} \cdot \mbox{\boldmath $\sigma$} \xi
\nonumber \\
&& \hspace{1.5in}
\; + \; 6 \, { 1 \over m_c^2} \,
    \xi'^\dagger q'^m \sigma^n \eta' \eta^\dagger 
     ( q^m \sigma^n + q^n \sigma^m ) \xi 
\;\bigg] \;.
\label{ggLO:2}
\end{eqnarray}
Dividing by $(2 m_c)^2$ to account for  the difference 
in the normalization of states, we reproduce twice the corresponding 
terms in ${\rm Im} {\cal M}$ given by the sum of (A9a) and (A9b) 
of Ref. \cite{B-B-L}.

The terms in (\ref{match-gg}) proportional to $n_f \alpha_s^3$ 
come from the renormalization of the coupling constant in (\ref{ggLO:1})
and from the cuts of the Feynman diagrams in Figure 2 that pass
through 2 gluon lines. The renormalization contribution in the 
$\overline{MS}$ prescription is obtained by multiplying (\ref{ggLO:1})
by $Z_g^4$, where
\begin{equation}
Z_g \;=\; 1 \; + \; { n_f \over 12 \pi } \, \alpha_s \, 
       \left( { 1 \over \epsilon_{UV} } + \ln ( 4 \pi ) - \gamma \right ) \;.
\label{Z_g}
\end{equation}
The subscript $UV$ on the pole in $\epsilon$ is a reminder that it
has an ultraviolet origin.
In the diagrams obtained by cutting two gluon lines in Figure 2, the quark 
loops have ultraviolet divergences and collinear infrared divergences. 
With dimensional regularization, there is a calcellation between the 
ultraviolet and infrared divergences and the diagrams vanish. 
Making this cancellation
explicit, the sum of the diagrams can be expressed as  (\ref{ggLO:1}) 
multiplied by the factor
\begin{equation}
- {n_f \over 3 \pi } \; \alpha_s \; 
  \left(  {1 \over \epsilon_{UV} } \; - \; 
          {1 \over \epsilon_{IR} }  \right)   \;.
\end{equation}
The ultraviolet pole in $\epsilon$ is cancelled by the renormalization 
factor (\ref{Z_g}).
Adding the two contributions proportional to $n_f \alpha_s^3$, we obtain 
\begin{eqnarray}
&&
n_f \;
{8 \Gamma({3 \over 2}) 
	\over27  N (N+2) \Gamma({N \over 2})} \;
 \alpha_s^3 \mu^{4 \epsilon}
\left( {m_c^2 \over 4 \pi} \right)^{- \epsilon} \;
\left( { 1 \over \epsilon_{IR} } + \ln ( 4 \pi ) - \gamma \right )
\nonumber \\
& & \hspace{1.0in}
\times \;\Bigg[\; 
\; 6 (N+2) \,  
\left( \mbox{$-{1 \over 96}$}
	\xi'^\dagger \{ [ \sigma^r, \sigma^s ], \sigma^t \} \eta' 
	\eta^\dagger \{ [ \sigma^r, \sigma^s ], \sigma^t \} \xi \right)
\nonumber \\
& & \hspace{1.5in}
\; + \; (5N+7) \, { 1 \over m_c^2} \,
    \xi'^\dagger {\bf q}'\cdot \mbox{\boldmath $\sigma$} \eta' 
    \eta^\dagger {\bf q} \cdot \mbox{\boldmath $\sigma$} \xi
\nonumber \\
& & \hspace{1.5in}
\; + \; (2 N^2 + N - 9) \, { 1 \over m_c^2} \,
\xi'^\dagger q'^m \sigma^n \eta' 
   	\eta^\dagger (q^m \sigma^n +  q^n \sigma^m ) \xi 
\;\Bigg] \; .
\label{ggNLO:1}
\end{eqnarray}

On the QCD side of the matching condition, the total contribution 
proportional to $n_f \alpha_s^3$ is the sum of (\ref{gqqbar:3})
and (\ref{ggNLO:1}).  The pole in $\epsilon$ 
in front of the spinor factor 
$ \xi'^\dagger \{ [ \sigma^r, \sigma^s ], \sigma^t \} \eta' 
  \eta^\dagger \{ [ \sigma^r, \sigma^s ], \sigma^t \} \xi$ 
cancels in the sum.  The spinor factor can therefore be
simplified using (\ref{spinor-anti}). Adding 
(\ref{gqqbar:3}) and (\ref{ggNLO:1}) and keeping only the terms that 
survive in the limit $\epsilon \to 0$, we obtain
\begin{eqnarray}
&&
n_f \;
{8 \Gamma({3 \over 2}) \over 27 N (N+2) \Gamma({N \over 2})} \;
\alpha_s^3 \mu^{4 \epsilon}
\left( {m_c^2 \over 4 \pi} \right)^{- \epsilon} \;
\Bigg\{ \; 
- 60 \left( \ln {\mu \over 2 m_c} + {4 \over 3} \right) \,  
	\xi'^\dagger \eta' \eta^\dagger \xi
\nonumber \\
& & \hspace{1.0in}
\;-\; 44 \left( \ln {\mu \over 2 m_c} + {29 \over 33} \right) \,  
{ 1 \over m_c^2} \,
    \xi'^\dagger {\bf q}'\cdot \mbox{\boldmath $\sigma$} \eta' 
    \eta^\dagger {\bf q} \cdot \mbox{\boldmath $\sigma$} \xi
\nonumber \\
& & \hspace{1in}
\; + \; \left[ 
	- 20 \left( {1 \over \epsilon_{IR} } + \ln(4 \pi) - \gamma \right)
	- 64 \left( \ln {\mu \over 2 m_c} + {7 \over 12} \right) \right] 
{ 1 \over m_c^2} \,
	\xi'^\dagger q'^m \sigma^n \eta' \eta^\dagger q^m \sigma^n \xi 
\nonumber \\
& & \hspace{1in}
\;-\; 24 \left( \ln {\mu \over 2 m_c} + {1 \over 2} \right) \,  
{ 1 \over m_c^2} \,
	\xi'^\dagger q'^m \sigma^n \eta' \eta^\dagger q^n \sigma^m \xi 
\;\Bigg\} \;.
\label{match-QCD}
\end{eqnarray}

\subsection{ Short-distance coefficients }

We now calculate the matrix elements that must appear on  
the NRQCD side of the matching condition (\ref{match})
and extract their short-distance coefficients. 
We will find that the terms that are required for matching are
\begin{eqnarray}
&&
{C^{(\underline{1},^1S)} \over m_c^{N-1}} 
\langle c \bar c' | \psi^\dagger \chi \chi^\dagger \psi | c \bar c \rangle
\;+\; {C^{(\underline{8},^3S)} \over m_c^{N-1}} 
\langle c \bar c' | \psi^\dagger \mbox{\boldmath $\sigma$} T^a \chi 
	\cdot \chi^\dagger \mbox{\boldmath $\sigma$} T^a \psi 
	| c \bar c \rangle^{(\mu)}
\nonumber \\
& & \hspace{1in}
\;+\; {C^{(\underline{1},^3P)}_1 \over m_c^{N+1}}
\langle c \bar c' | \psi^\dagger (-\mbox{$\frac{i}{2}$} \tensor{\bf D} 
	\cdot \mbox{\boldmath $\sigma$}) \chi \chi^\dagger 
	(-\mbox{$\frac{i}{2}$} \tensor{\bf D} \cdot 
	\mbox{\boldmath $\sigma$}) \chi | c \bar c \rangle
\nonumber \\
& & \hspace{1in}
\;+\; {C^{(\underline{1},^3P)}_2(\mu) \over m_c^{N+1}}
\langle c \bar c' | \psi^\dagger ( -\mbox{$\frac{i}{2}$} \tensor{D})^m \sigma^n \chi \; 
	\chi^\dagger ( -\mbox{$\frac{i}{2}$} \tensor{D})^m \sigma^n \psi
	 | c \bar c \rangle
\nonumber \\
& & \hspace{1in}
\;+\; {C^{(\underline{1},^3P)}_3 \over m_c^{N+1}}
\langle c \bar c' | \psi^\dagger ( -\mbox{$\frac{i}{2}$} \tensor{D})^m 
	\sigma^n \chi \; \chi^\dagger ( -\mbox{$\frac{i}{2}$} \tensor{D})^n 
	\sigma^m \psi | c \bar c \rangle
\;+\; \ldots  \;.
\label{match-NRQCD}
\end{eqnarray}
We have placed a superscript $(\mu)$ on the color-octet matrix element
in anticipation of the fact that it will acquire dependence on the 
NRQCD renormalization scale through  renormalization.

We consider the matching condition first for color-octet $c \bar{c}$ pairs 
and then for color-singlet $c \bar{c}$ pairs.
In the case of color-octet $c \bar{c}$ pairs, the contribution
proportional to $n_f \alpha_s^2$ on the QCD side of 
the matching condition is given in (\ref{qqbar:2}).
The spinor factor in (\ref{qqbar:2})
can be identified as the 
expression at leading order in $\alpha_s$ and expanded to linear order in 
${\bf q}$ and ${\bf q}'$ of the following NRQCD matrix element:
\begin{equation}
\langle c\bar{c}'  | \psi^\dagger \mbox{\boldmath $\sigma$} T^a \chi \cdot 
	\chi^\dagger \mbox{\boldmath $\sigma$} T^a \psi | c\bar{c} \rangle 
\;\approx\; 4 m_c^2 \; 
	\xi'^\dagger \mbox{\boldmath $\sigma$} T^a \eta' \cdot
	\eta^\dagger \mbox{\boldmath $\sigma$} T^a \xi \;.
\label{matel-ij}
\end{equation}
The tree-level expression for the matrix element in (\ref{matel-ij})
is represented diagramatically in Fig. 4, with 
the dot representing the operator.
Comparing (\ref{qqbar:2}) and (\ref{match-NRQCD}), we can read off
the short-distance coefficient of the matrix element in (\ref{matel-ij}):
\begin{equation}
C^{(\underline{8},^3S)} 
\;=\; n_f \; 
{\pi (N-1) \Gamma({3 \over 2}) \over 2 N \Gamma({N \over 2})}
\alpha_s^2
\left( {4 \pi \mu^4 \over m_c^4} \right)^\epsilon \;.
\label{C-83S:N}
\end{equation}
Taking the limit $N \to 3$, this reduces to 
\begin{equation}
C^{(\underline{8},^3S)} 
\;=\;  {\pi n_f \over 3} \alpha_s^2(2 m_c)  \;.
\label{C-83S:3}
\end{equation}
Since the short-distance coefficient is sensitive only to momenta 
of order $m_c$, we have set the scale of the running coupling constant 
to $\mu = 2 m_c$.

In the case of color-singlet $c \bar{c}$ pairs, the contributions 
proportional to $\alpha_s^2$ and $n_f \alpha_s^3$ on the QCD side 
of the matching condition are given in (\ref{ggLO:2}) and 
(\ref{match-QCD}), respectively.
The spinor factors in these expressions can be identified as  
tree-level expressions for NRQCD matrix elements,
expanded to linear order in ${\bf q}$ and ${\bf q}'$.
The S-wave spinor factor is proportional to 
\begin{equation}
\langle c\bar{c}' | \psi^\dagger \chi \; 
	\chi^\dagger \psi | c\bar{c} \rangle
\; \approx \; 4 m_c^2 \;
\xi'^\dagger \eta' \eta^\dagger \xi \;.
\label{matel-singS}
\end{equation}
Comparing the sum of (\ref{ggLO:2}) and 
(\ref{match-QCD}) with (\ref{match-NRQCD}), 
we can read off the short-distance coefficient of the matrix element 
in (\ref{matel-singS}):
\begin{equation}
C^{(\underline{1},^1S)} 
\;=\;
{4 \pi \over 9} \alpha_s^2(\mu)
\left[ 1 \;+\;  \left(- {2 \over 3} \ln{\mu \over 2 m_c} - {8 \over 9} \right)
	n_f {\alpha_s \over \pi} \right] \;.
\label{C-13S:mu}
\end{equation}
Choosing the QCD renormalization scale to be  $\mu = 2 m_c$, 
this reduces to 
\begin{equation}
C^{(\underline{1},^1S)} 
\;=\; {4 \pi \over 9} \alpha_s^2(2 m_c)
 \left( 1 \;-\; {8 n_f \over 9} {\alpha_s \over \pi} \right)  \;.
\label{C-13S}
\end{equation}

The P-wave spinor factors in (\ref{ggLO:2}) and (\ref{match-QCD})
can be identified with the matrix elements
\begin{mathletters}
\label{matel-singP}
\begin{eqnarray}
\langle c\bar{c}' |  \psi^\dagger (-\mbox{$\frac{i}{2}$} \tensor{\bf D} 
		\cdot \mbox{\boldmath $\sigma$}) \chi \; 
\chi^\dagger (-\mbox{$\frac{i}{2}$} \tensor{\bf D} 
		\cdot \mbox{\boldmath $\sigma$}) \psi | c\bar{c}  \rangle 
& \approx & 4 m_c^2 \;
\xi'^\dagger {\bf q}' \cdot \mbox{\boldmath $\sigma$} \eta' 
	\eta^\dagger {\bf q} \cdot \mbox{\boldmath $\sigma$} \xi ,
\label{matel-singP:1}
\\
\langle  c\bar{c}' | 
	\psi^\dagger ( -\mbox{$\frac{i}{2}$} \tensor{D})^m \sigma^n \chi \; 
	\chi^\dagger ( -\mbox{$\frac{i}{2}$} \tensor{D})^m \sigma^n \psi 
	| c\bar{c}  \rangle
&\approx & 4 m_c^2 \;
\xi'^\dagger q'^m \sigma^n \eta' \eta^\dagger q^m \sigma^n \xi ,
\label{matel-singP:2}
\\
\langle c\bar{c}' |  
\psi^\dagger ( -\mbox{$\frac{i}{2}$} \tensor{D})^m \sigma^n \chi \; 
	\chi^\dagger ( -\mbox{$\frac{i}{2}$} \tensor{D})^n \sigma^m \psi 
	| c\bar{c}  \rangle
& \approx & 4 m_c^2 \;
\xi'^\dagger q'^m \sigma^n \eta' \eta^\dagger q^n \sigma^m \xi \;.
\label{matel-singP:3}
\end{eqnarray}
\end{mathletters}
For the matrix elements (\ref{matel-singP:1})
and (\ref{matel-singP:3}), we can immediately read off 
the short-distance coefficients by comparing the sum of 
(\ref{ggLO:2}) and (\ref{match-QCD}) with (\ref{match-NRQCD}).
Setting $N=3$ and $\mu = 2 m_c$, we obtain
\begin{eqnarray}
C^{(\underline{1},^3P)}_1 
&=& {44 \pi \over 135} \alpha_s^2(2 m_c)
 \left( 1 \;-\; {58 n_f \over 99} {\alpha_s \over \pi} \right)  \;,
\label{C-13P:1}
\\
C^{(\underline{1},^3P)}_3 
&=& {8 \pi \over 45} \alpha_s^2(2 m_c)
 \left( 1 \;-\; {n_f \over 3} {\alpha_s \over \pi} \right)  \;.
\label{C-13P:3}
\end{eqnarray}

The coefficient of the matrix element  in (\ref{matel-singP:1})
is more complicated because the coefficient of 
$\xi'^\dagger q'^m \sigma^n \eta' \eta^\dagger q^m \sigma^n \xi$
in (\ref{match-QCD}) contains an infrared pole in $\epsilon$,
indicating that it is sensitive to long-distance effects.
Since an infrared divergence cannot appear in a short-distance coefficient,
that divergence must be matched by an infrared divergence 
in a radiative correction to a
matrix element on the NRQCD side of the matching condition.
Since the divergence
in (\ref{match-QCD}) has a coefficient proportional to  $n_f \alpha_s^3$,
the infrared-divergent NRQCD matrix element  must have a 
short-distance coefficient  proportional to $n_f \alpha_s^2$.  The only such 
matrix element is 
$\langle c \bar c' | \psi^\dagger \mbox{\boldmath $\sigma$} T^a \chi \cdot
	\chi^\dagger \mbox{\boldmath $\sigma$} T^a \psi | c \bar c \rangle$,
whose short-distance coefficient has already been 
determined in (\ref{C-83S:N}).  
Thus the infrared divergence on the NRQCD
side of the matching condition must come from that term.

If the initial and final
$c \bar c$ pairs are in color-singlet states, the tree level
expression (\ref{matel-ij}) for the matrix element vanishes
and the leading contribution comes instead from radiative corrections.
The leading contributions to the matrix element are represented by the 
diagrams in Figure 5.  The expression for diagram 5a is
\begin{eqnarray}
& & 
4 g^2 \mu^{2 \epsilon} \;
\xi'^\dagger \sigma^n T^a T^b \eta' 
	\eta^\dagger \sigma^n T^b T^a \xi 
\int {d^Nk \over (2 \pi)^N 2k}
\left( {\bf q} \cdot {\bf q}' 
	- {\bf q} \cdot \hat{\bf k} \; \hat{\bf k} \cdot {\bf q}' \right)
\nonumber \\
& & \hspace{1in} \times
{ 1 \over E_q - k - ({\bf q} - {\bf k})^2/(2 m_c) + i \epsilon} \;
{ 1 \over E_q - k - ({\bf q}' - {\bf k})^2/(2 m_c) + i \epsilon} \;,
\label{fig-5a}
\end{eqnarray}
where $E_q = {\bf q}^2/(2 m_c) = ({\bf q}')^2/(2 m_c)$.  
Since the initial and final $c\bar{c}$ pairs are both color-singlets,
we can replace $T^a T^b$ and $T^b T^a$ by $\delta^{ab}/6$. 
The proper way to evaluate the diagram is to regard 
$|{\bf q}|$, $|{\bf q}'|$, and $|{\bf k}|$ to all be much smaller than $m_c$.
We must therefore expand out the denominators  of (\ref{fig-5a})
in powers of ${\bf q}/m_c$, ${\bf q}'/m_c$, and ${\bf k}/m_c$
before integrating over ${\bf k}$.  Keeping only terms up to 
linear order in ${\bf q}/m_c$ and ${\bf q}'/m_c$, the diagram reduces to 
\begin{equation}
{16 \pi \over 9} \alpha_s \mu^{2 \epsilon} \;
\xi'^\dagger \sigma^n  \eta' \eta^\dagger \sigma^n  \xi 
\int {d^Nk \over (2 \pi)^N}
{ {\bf q} \cdot {\bf q}' 
	- {\bf q} \cdot \hat{\bf k} \; \hat{\bf k} \cdot {\bf q}' 
	\over k^3 } \;.
\label{int-k}
\end{equation}
The integral is both ultraviolet and infrared divergent.  It vanishes in 
dimensional regularization due to a cancellation between an
ultraviolet pole in $\epsilon$ and an infrared pole.  
Making these poles explicit, the diagram can be written
\begin{equation}
{8  \over 27 \pi} \, \alpha_s \, 
\left( {1 \over \epsilon_{UV}} - {1 \over \epsilon_{IR}} \right)
\xi'^\dagger {q'}^m \sigma^n \eta' 
	\eta^\dagger q^m \sigma^n  \xi \;.
\end{equation}
The subscripts $UV$ and $IR$ on $\epsilon$
indicate whether the pole is of ultraviolet or infrared origin.
We have set $N=3$ in the prefactor, since any finite terms obtained by 
expanding the prefactor in powers of $\epsilon$ will cancel. 
The  other 3 diagrams in Figure 5 give identical
contributions.   The final result for the 
matrix element is 
\begin{equation} 
\langle c\bar{c}' | \psi^\dagger \mbox{\boldmath $\sigma$} T^a \chi \cdot 
        \chi^\dagger \mbox{\boldmath $\sigma$} T^a \psi | c\bar{c} \rangle 
\;=\; {32 \alpha_s \over 27 \pi} 
\left( {1 \over \epsilon_{UV}} - {1 \over \epsilon_{IR}} \right)
\xi'^\dagger {q'}^m \sigma^n \eta' 
	\eta^\dagger q^m \sigma^n \xi \;.
\label{matel-bare}
\end{equation}
After multiplying by $C^{(\underline{8},^1S)}/m_c^{N-1}$, where 
$C^{(\underline{8},^1S)}$ is given in (\ref{C-83S:N}), 
we find that the infrared pole in $\epsilon$
matches the one on the QCD side of the matching condition, 
which is given in (\ref{match-QCD}).

The ultraviolet pole in $\epsilon$ in  (\ref{matel-bare}) indicates 
that the matrix element is ultraviolet divergent and therefore 
requires renormalization.  It is the renormalized matrix element
that appears in the NRQCD side of the matching condition (\ref{match-NRQCD}).
In the $\overline{MS}$ renormalization scheme,  
the relation between matrix elements of the bare operator 
and renormalized operators is 
\begin{eqnarray}
&&
\langle c\bar{c}' | \psi^\dagger \mbox{\boldmath $\sigma$} T^a \chi \cdot 
        \chi^\dagger \mbox{\boldmath $\sigma$} T^a \psi | c\bar{c}  \rangle
\;=\; \mu^{- 4 \epsilon} \Bigg(
\langle c\bar{c}' | \psi^\dagger \mbox{\boldmath $\sigma$} T^a \chi \cdot 
        \chi^\dagger \mbox{\boldmath $\sigma$} T^a \psi 
        | c\bar{c} \rangle^{(\mu)}  
\nonumber \\
 & & \hspace{0.5in}
\;+\; {8 \alpha_s \over 27 \pi m_c^2} 
\left( {1 \over \epsilon_{UV}} + \ln(4 \pi) - \gamma \right)
\langle c\bar{c}' | 
\psi^\dagger ( -\mbox{$\frac{i}{2}$} \tensor{D})^m \sigma^n \chi \, 
\chi^\dagger ( -\mbox{$\frac{i}{2}$} \tensor{D})^m 
	\sigma^n \psi | c\bar{c} \rangle^{(\mu)} \Bigg) \; .
\label{matel-ren:def}
\end{eqnarray}
The superscripts $(\mu)$ on the matrix elements on the right
side indicate that they are renormalized matrix elements with 
renormalization scale $\mu$.  
We will suppress this superscript on color-singlet matrix elements,
since they do not require any renormalization at this order in $\alpha_s$.
The  fermion field operators in the bare matrix element on the left side 
of (\ref{matel-ren:def}) have dimension $N/2$.  
The fermion field operators in the renormalized 
matrix elements on the right side have dimension $3/2$.
The factor of $\mu^{-4 \epsilon}$ on the right side of (\ref{matel-ren:def})
compensates for the difference between the dimensions of the two sides. 
Solving (\ref{matel-ren:def}) for 
$\langle c\bar{c}' | \psi^\dagger \mbox{\boldmath $\sigma$} T^a \chi \cdot 
	\chi^\dagger \mbox{\boldmath $\sigma$} T^a | c\bar{c} \rangle^{(\mu)}$ 
and using (\ref{matel-bare}) and (\ref{matel-singP:2}),        
we find that the renormalized matrix element is
\begin{equation} 
\langle c\bar{c}' | \psi^\dagger \mbox{\boldmath $\sigma$} T^a \chi \cdot 
        \chi^\dagger \mbox{\boldmath $\sigma$} T^a \psi 
        | c\bar{c} \rangle^{(\mu)}
\;=\; - {32  \over 27 \pi} \alpha_s 
\left( {1 \over \epsilon_{IR}}  + \ln(4 \pi) - \gamma \right)
\xi'^\dagger {q'}^m \sigma^n \eta' 
	\eta^\dagger q^m \sigma^n \xi \;.
\label{matel-ren}
\end{equation}

Multiplying by $C^{(\underline{8},^1S)}/m_c^{N-1}$, where 
$C^{(\underline{8},^1S)}$ is given in (\ref{C-83S:N})
and keeping all terms that survive in the
limit $\epsilon \to 0$, we find that the contribution to the 
NRQCD side of the matching condition is
\begin{eqnarray}
&&
{C^{(\underline{8},^3S)} \over  m_c^{N-1}} \; 
\langle c \bar{c}'  | \psi^\dagger \mbox{\boldmath $\sigma$} T^a \chi \cdot
        \chi^\dagger \mbox{\boldmath $\sigma$} T^a \psi 
        |  c \bar{c}  \rangle^{(\mu)}
\;=\;  n_f 
{8 \Gamma({3 \over 2}) \over 27 N (N+2) \Gamma({N \over 2}) }
\alpha_s^3 \mu^{4 \epsilon} 
\left( {m_c^2 \over 4 \pi} \right)^{- \epsilon}
\nonumber \\
&& \hspace{1.5in}
\times \left[ - 20
\left( {1 \over \epsilon_{IR}} + \ln(4 \pi) 
	- \gamma - {7 \over 5} \right) \right] \;
{1 \over m_c^2} \xi'^\dagger {q'}^m \sigma^n \eta' 
	\eta^\dagger q^m \sigma^n \xi \;.
\label{match-NRQCD:NLO}
\end{eqnarray}
We have expressed this in a form that makes it as easy as possible to 
match with (\ref{match-QCD}), which is the term proportional to 
$n_f \alpha_s^3$ on the QCD side of the matching condition.
In particular, we have expanded out a factor of $(N-1)(N+2)$
to get the constant $-{7 \over 5}$ under the pole in $\epsilon$.
We see that the infrared poles in $\epsilon$ in (\ref{match-QCD})
and  (\ref{match-NRQCD:NLO}) match.
The remainder of that term must be matched by the term proportional to
$C^{(\underline{1},^3P)}_2$ in (\ref{match-NRQCD}).  The resulting
expression for the coefficient is 
\begin{equation}
C^{(\underline{1},^3P)}_2 
\;=\;
{8 \pi \over 45} \alpha_s^2(\mu)
\left[ 1 \;+\; 
	\left( - {16 \over 9} \ln{\mu \over 2 m_c} - {49 \over 27} \right)
	{\alpha_s \over \pi} \right] \;.
\label{C-13P:2mu}
\end{equation}
Shifting the renormalization scale of the QCD coupling constant to
$\mu = 2 m_c$ we obtain 
our final result:
\begin{equation}
C^{(\underline{1},^3P)}_2 (\mu)
\;=\;
{8 \pi \over 45} \alpha_s^2(2 m_c)
\left[ 1 \;+\; 
	\left( - {10 \over 9} \ln{\mu \over 2 m_c} - {49 \over 27} \right)
	{\alpha_s \over \pi} \right] \;.
\label{C-13P:2}
\end{equation}
The remaining logarithms of $\mu$ represent the dependence of the 
matrix element on the renormalization scale $\mu$ of NRQCD.

\section{ Annihilation Decay Rates }

The factorization formula (\ref{fact-Gam}) for the annihilation decay 
rate of a quarkonium state $H$, including all terms whose coefficients 
have been computed explicitly in Section III, is
\begin{eqnarray}
\Gamma(H) 
& = &  {1 \over 2 M_H } \Bigg( 
{C^{(\underline{1},^1S)} \over m_c^2} \; 
\langle H | \psi^\dagger \chi \; 
	\chi^\dagger \psi | H \rangle 
\;+\; {C^{(\underline{8},^3S)} \over m_c^2} \; 
\langle H | \psi^\dagger \mbox{\boldmath $\sigma$} T^a \chi \cdot 
	  \chi^\dagger \mbox{\boldmath $\sigma$} T^a \psi | H \rangle^{(\mu)}
\nonumber \\
 & & \hspace{1in}
\; + \; {C^{(\underline{1},^3P)}_1 \over m_c^4} \; 
\langle H | \psi^\dagger (-\mbox{$\frac{i}{2}$} \tensor{\bf D} 
		\cdot \mbox{\boldmath $\sigma$}) \chi \; 
\chi^\dagger (-\mbox{$\frac{i}{2}$} \tensor{\bf D} 
		\cdot \mbox{\boldmath $\sigma$}) \psi | H \rangle 
\nonumber \\
 & & \hspace{1in}
\; + \; {C^{(\underline{1},^3P)}_2(\mu) \over m_c^4} \; 
\langle H | \psi^\dagger (-\mbox{$\frac{i}{2}$} \tensor{D})^m \sigma^n 
	\chi \; \chi^\dagger (-\mbox{$\frac{i}{2}$} \tensor{D})^m 
	\sigma^n \psi | H \rangle
\nonumber \\
 & & \hspace{1in}
\; + \; {C^{(\underline{1},^3P)}_3 \over m_c^4} \; 
\langle H | \psi^\dagger (-\mbox{$\frac{i}{2}$} \tensor{D})^m 
	\sigma^n \chi \; \chi^\dagger (-\mbox{$\frac{i}{2}$} \tensor{D})^n 
	\sigma^m \psi | H \rangle 
\;+\; \ldots \Bigg) \; .
\label{fact-Gam:3}
\end{eqnarray}
We have calculated the terms proportional to $n_f \alpha_s^2$
in $C^{(\underline{8},^3S)}$ and the terms proportional to $\alpha_s^2$
or $n_f \alpha_s^3$ in the other coefficients in (\ref{fact-Gam:3}).
The $\mu$-dependence of the coefficient $C^{(\underline{1},^3P)}_2$
cancels that of the renormalized matrix element, which satisfies
\begin{eqnarray}
&&
\mu {d \ \over d \mu} 
\langle H | \psi^\dagger \mbox{\boldmath $\sigma$} T^a \chi \cdot 
	  \chi^\dagger \mbox{\boldmath $\sigma$} T^a \psi | H \rangle^{(\mu)}
\nonumber \\
&& \hspace{1in}
\;=\; {16 \over 27 \pi} {\alpha_s(\mu) \over m_c^2}
\langle H | \psi^\dagger (-\mbox{$\frac{i}{2}$} \tensor{D})^m \sigma^n 
	\chi \; \chi^\dagger (-\mbox{$\frac{i}{2}$} \tensor{D})^m 
	\sigma^n \psi | H \rangle  \; .
\label{matel-rg}
\end{eqnarray}
The relative importance of the various terms in 
the annihilation decay rates (\ref{fact-Gam:3})
depends on the quarkonium state.  The magnitude of a particular term
is determined by the order in $\alpha_s$ of its short-distance coefficient
and by the scaling of the matrix element with $v$,
which is given by the velocity-scaling rules of NRQCD \cite{B-B-L}. 
Below, we apply this general formula to
the annihilation decay rates of spin-singlet S-wave states 
and spin-triplet P-wave states.

\subsection{Spin-singlet S-wave states}

The dominant Fock state of the $\eta_c$ consists of a $c \bar c$ pair in a 
color-singlet ${}^1S_0$ state.  
The largest matrix element that appears in the NRQCD factorization formula 
for the decay rate is therefore 
$\langle \eta_c | \psi^\dagger \chi \chi^\dagger \psi | \eta_c \rangle$, 
which scales as $v^3$. 
The next most important matrix element is
$\langle \eta_c | (\psi^\dagger \chi \chi^\dagger {\bf D}^2 \psi 
	+ {\rm h.c.}) | \eta_c \rangle$, 
which is suppressed by a factor of $v^2$. 
Matrix elements whose dominant
contributions come from higher Fock states are suppressed by $v^3$ or more.
For example, the matrix element whose dominant contribution comes from 
the $c \bar c g$ Fock state in which the $c \bar c$ pair 
is in a color-octet $^1P_1$ state 
is suppressed by $v^4$, with one factor of $v^2$ arising from 
the probability of the $c \bar c g$ Fock state
and  the other arising from the derivatives in the P-wave operator.
The matrix element 
$\langle \eta_c | \psi^\dagger \mbox{\boldmath $\sigma$} T^a \chi \cdot 
	\chi^\dagger \mbox{\boldmath $\sigma$} T^a \psi | \eta_c \rangle$,
whose dominant contribution comes from 
the $c \bar c g$ Fock state in which the $c \bar c$ pair 
is in a color-octet $^3S_1$ state, is suppressed by $v^3$ from the 
probability of that $c \bar c g$ Fock state.  
Thus, up to corrections of relative order $v^2$,
the  annihilation decay rate can be written
\begin{equation}
\Gamma (\eta_c) \;= \; {1 \over 2 M_{\eta_c} }
{C^{(\underline{1},^1S)} \over m_c^2}
\langle \eta_c | \psi^\dagger  \chi  \,
        \chi^\dagger  \psi  | \eta_c  \rangle \;.
\label{wid-etac}
\end{equation}
The short-distance coefficient, including the next-to-leading order correction 
proportional to $n_f \alpha_s^3$, is given in (\ref{C-13S}).
The $n_f \alpha_s^3$ term agrees with the complete 
next-to-leading order correction calculated by
Barbieri et al. \cite{Barbieri:etac} 
and by Hagiwara et al. \cite{Hagiwara}. 

The difference between the matrix element in (\ref{wid-etac})
and the standard NRQCD matrix element 
$\langle \eta_c | {\cal O}_1(^{1}S_0) | \eta_c  \rangle$ 
introduced in Ref. \cite{B-B-L}
is discussed in Appendix B of \cite{Braaten-Chen}.  Up to corrections 
of relative order $v^2$, the difference is simply an overall 
normalization factor:
\begin{equation}
\langle \eta_c | \psi^\dagger \chi \, 
        \chi^\dagger \psi | \eta_c \rangle 
\; \approx \;
4 m_c \; \langle \eta_c | {\cal O}_1 ({}^1S_0) | {\eta_c} \rangle \;.
\end{equation}
Using the vacuum-saturation approximation, this matrix element
can be related to the radial wavefunction  of the $\eta_c$
evaluated at the origin:
\begin{equation}
\langle \eta_c | {\cal O}_1 ({}^1S_0) | {\eta_c} \rangle
\; \approx \; 
{3 \over 2 \pi} |R_{\eta_c}(0)|^2 \;.
\end{equation}
This approximation is accurate up to corrections of relative order $v^4$.

\subsection{Spin-triplet P-wave states}

The dominant Fock state of the $\chi_{cJ}$ consists of a $c \bar c$ pair 
in a color-singlet ${}^3P_J$ state.  This Fock state gives the dominant 
contributions to  the color-singlet P-wave matrix elements in 
(\ref{fact-Gam:3}).
Since the derivatives in the P-wave operators give a suppression by 
$v^2$, the color-singlet P-wave matrix elements scale  as $v^5$.  
The matrix element
$\langle \chi_{cJ} |\psi^\dagger \mbox{\boldmath $\sigma$} T^a \chi \cdot 
	\chi^\dagger \mbox{\boldmath $\sigma$} T^a \psi | \chi_{cJ} \rangle$,
whose dominant contribution comes from 
the $c \bar c g$ Fock state in which the $c \bar c$ pair 
is in a color-octet $^3S_1$ state, is suppressed by a factor of $v^2$ 
from the probability of that $c \bar c g$ Fock State.
Thus, it also scales like $v^5$, like the color-singlet P-wave matrix 
elements.  All other matrix elements are suppressed by $v^2$ or more.
Thus, up to corrections of relative order $v^2$,
the decay rate is given by the sum of the 
color-octet S-wave term and the three color-singlet P-wave terms
in the NRQCD factorization formula (\ref{fact-Gam:3}).

We will use rotational symmetry and the approximate heavy-quark spin symmetry 
of NRQCD to show that the  color-octet S-wave matrix elements
$\langle \chi_{cJ} |\psi^\dagger \mbox{\boldmath $\sigma$} T^a \chi \cdot 
	\chi^\dagger \mbox{\boldmath $\sigma$} T^a \psi | \chi_{cJ} \rangle$
for the three $\chi_{cJ}$ states
are equal up to corrections of relative order $v^2$.  
Spin symmetry  implies that the state
$\chi_{cJ}(j_z)$, which is an eigenstate of ${\bf J}^2$ and $J_z$,
can be expressed in the form
\begin{equation}
\Big | \chi_{cJ}(j_z) \Big \rangle 
\;\approx\; \sum_{l_z s_z} \langle 1 l_z; 1 s_z | J j_z \rangle
\Big | \chi_c(l_z s_z) \Big \rangle \;,
\label{chi}
\end{equation}
where  $\chi_c(l_z s_z)$ is an eigenstate of $L_z$ and $S_z$.
Thus the color-octet $^3S_1$ matrix element can be written
\begin{eqnarray}
\langle \chi_{cJ} | \psi^\dagger \mbox{\boldmath $\sigma$} T^a \chi \cdot
	\chi^\dagger \mbox{\boldmath $\sigma$} T^a \psi  | \chi_{cJ}\rangle 
&\equiv& 
{1 \over 2 J + 1} \; \sum_{j_z}
\langle \chi_{cJ}(j_z) | \psi^\dagger \mbox{\boldmath $\sigma$} T^a \chi 
	\cdot \chi^\dagger \mbox{\boldmath $\sigma$} T^a \psi 
	| \chi_{cJ}(j_z) \rangle 
\nonumber \\
&\approx& 
{1 \over  2 J + 1} \sum_{j_z}
\sum_{l_z' s_z'} \sum_{l_z s_z} 
	\langle J j_z | 1 l_z'; 1 s_z' \rangle 
	\langle 1 l_z; 1 s_z | J j_z \rangle 
\nonumber \\
&& \hspace{1in} \times 
\Big \langle \chi_{c}(l_z' s_z') \Big | \psi^\dagger \sigma^i T^a \chi 
	\chi^\dagger \sigma^i T^a \psi \Big | \chi_{c}(l_z s_z) \Big \rangle \;.
\label{octS-SS}
\end{eqnarray}
Spin symmetry also implies that the matrix element on the right side
of  (\ref{octS-SS}) is proportional to $U^\dagger_{i s_z'} U_{s_z i}$, where
$U_{mi}$ is the unitary $3 \times 3$ matrix that transforms vectors from the 
Cartesian basis to the spherical basis.
Finally, rotational symmetry implies that matrix elements on the right side
of (\ref{octS-SS}) must be proportional to $\delta_{l_z' l_z}$.
Using the orthogonality relations of the Clebsch-Gordan coefficients,
the equation (\ref{octS-SS}) can be reduced to 
\begin{equation}
\langle \chi_{cJ} | \psi^\dagger \mbox{\boldmath $\sigma$} T^a \chi \cdot 
	\chi^\dagger \mbox{\boldmath $\sigma$} T^a \psi  | \chi_{cJ}\rangle 
\;\approx\;  
\langle \chi_{c0} | \psi^\dagger \mbox{\boldmath $\sigma$} T^a \chi \cdot 
	\chi^\dagger \mbox{\boldmath $\sigma$} T^a \psi  | \chi_{c0}\rangle \;.
\label{octS-id}
\end{equation}
These relations hold up to corrections of relative order $v^2$.

We next show that the color-singlet P-wave matrix elements in (\ref{fact-Gam:3}) 
can be reduced to a single independent matrix element, which we choose to be
$\langle \chi_{c0} | \psi^\dagger (-\mbox{$\frac{i}{2}$} \tensor{\bf D} 
		\cdot \mbox{\boldmath $\sigma$}) \chi 
\chi^\dagger (-\mbox{$\frac{i}{2}$} \tensor{\bf D} 
		\cdot \mbox{\boldmath $\sigma$}) \psi | \chi_{c0}  \rangle$.
Using the expression (\ref{chi}) for the 
$\chi_{cJ}$
states, we can write
\begin{eqnarray}
\langle \chi_{cJ} | 
\psi^\dagger (-\mbox{$\frac{i}{2}$} \tensor{D})^m \sigma^n \chi 
\chi^\dagger (-\mbox{$\frac{i}{2}$} \tensor{D})^i \sigma^j \psi 
| \chi_{cJ} \rangle
\;\approx\; 
{1 \over  2 J + 1} \sum_{j_z}
\sum_{l_z' s_z'} \sum_{l_z s_z} 
	\langle J j_z | 1 l_z'; 1 s_z' \rangle 
	\langle 1 l_z; 1 s_z | J j_z \rangle 
\nonumber \\
\times  \Big \langle \chi_c(l_z' s_z')  \Big | 
	\psi^\dagger (-\mbox{$\frac{i}{2}$} \tensor{D})^m \sigma^n \chi
	\chi^\dagger (-\mbox{$\frac{i}{2}$} \tensor{D})^i \sigma^j \psi
	\Big |  \chi_c(l_z s_z)\Big \rangle \;.
\label{singP-SS}
\end{eqnarray}
The vacuum-saturation approximation, which is accurate up to corrections 
of relative order $v^4$ can be used to express the matrix element 
on the right side of (\ref{singP-SS}) as the  product of 
$\langle \chi_c(l_z' s_z') |
	\chi^\dagger (-\mbox{$\frac{i}{2}$} \tensor{D})^m \sigma^n \psi
	| 0 \rangle$ and
$\langle 0 |
	\psi^\dagger (-\mbox{$\frac{i}{2}$} \tensor{D})^i \sigma^j \chi
	| \chi_c(l_z s_z) \rangle$. 
Rotational symmetry and spin symmetry imply that these two matrix elements 
are proportional to $U^\dagger_{m l_z'} U^\dagger_{n s_z'}$ and 
$U_{l_z i} U_{s_z j}$, respectively.
Thus the tensorial structure
of the matrix element (\ref{singP-SS}) is completely determined.
The proportionality constant can be deduced by taking the special case 
$J=0$, $i=j$ and $m=n$, summed over $i$ and $m$.  The resulting formula is
\begin{eqnarray}
\langle \chi_{cJ} | 
\psi^\dagger (-\mbox{$\frac{i}{2}$} \tensor{D})^i \sigma^j \chi 
\chi^\dagger (-\mbox{$\frac{i}{2}$} \tensor{D})^m \sigma^n \psi 
| \chi_{cJ} \rangle
\;\approx\; {1 \over 3} \;
\langle \chi_{c0} |
\psi^\dagger  (-\mbox{$\frac{i}{2}$} \tensor{\bf D} 
		\cdot \mbox{\boldmath $\sigma$}) \chi 
\chi^\dagger  (-\mbox{$\frac{i}{2}$} \tensor{\bf D} 
		\cdot \mbox{\boldmath $\sigma$}) \psi 
| \chi_{c0}\rangle
\nonumber \\
\;\times\; 
{1 \over  2 J + 1} \sum_{j_z}
\sum_{l_z' s_z'} \sum_{l_z s_z} 
	\langle J j_z | 1 l_z'; 1 s_z' \rangle 
	\langle 1 l_z; 1 s_z | J j_z \rangle 
	U^\dagger_{m l_z'} U^\dagger_{n s_z'} U_{l_z i} U_{s_z j} \;.
\label{singP-tensor}
\end{eqnarray}
The scalar combinations of these matrix elements can be simplified
by using the identity 
$(U U^t)_{m_1 m_2} = - \sqrt{3} \langle 1 m_1; 1 m_2 | 0 0 \rangle$
together with the orthogonality relations for Clebsch-Gordan coefficients.
The resulting formulas are
\begin{mathletters}
\label{singP-scalar}
\begin{eqnarray}
 & & \langle \chi_{cJ} | \psi^\dagger (-\mbox{$\frac{i}{2}$} \tensor{\bf D} 
		\cdot \mbox{\boldmath $\sigma$}) \chi 
\chi^\dagger (-\mbox{$\frac{i}{2}$} \tensor{\bf D} 
		\cdot \mbox{\boldmath $\sigma$}) \psi | \chi_{cJ} \rangle
\nonumber \\
& &  \hspace{1.5in}
\;\approx\;  \delta_{J0} \;
\langle \chi_{c0} | \psi^\dagger  (-\mbox{$\frac{i}{2}$} \tensor{\bf D} 
		\cdot \mbox{\boldmath $\sigma$}) \chi 
\chi^\dagger  (-\mbox{$\frac{i}{2}$} \tensor{\bf D} 
		\cdot \mbox{\boldmath $\sigma$}) \psi | \chi_{c0}  \rangle \,,
\\
&& 
\langle \chi_{cJ} | \psi^\dagger (-\mbox{$\frac{i}{2}$} \tensor{D})^m 
	\sigma^n \chi \chi^\dagger (-\mbox{$\frac{i}{2}$} \tensor{D})^m 
	\sigma^n \psi | \chi_{cJ} \rangle
\nonumber \\
& &  \hspace{1.5in}
\;\approx\;  {1 \over 3} \;
\langle  \chi_{c0} | \psi^\dagger  (-\mbox{$\frac{i}{2}$} \tensor{\bf D} 
		\cdot \mbox{\boldmath $\sigma$}) \chi 
\chi^\dagger  (-\mbox{$\frac{i}{2}$} \tensor{\bf D} 
		\cdot \mbox{\boldmath $\sigma$}) \psi | \chi_{c0} \rangle \,,
\\
&& \langle \chi_{cJ} | \psi^\dagger (-\mbox{$\frac{i}{2}$} \tensor{D})^m 
	\sigma^n \chi \chi^\dagger (-\mbox{$\frac{i}{2}$} \tensor{D})^n 
	\sigma^m \psi | \chi_{cJ}\rangle
\nonumber \\
& &  \hspace{1.5in}
 \;\approx\; (-1)^J {1 \over 3} \;
\langle \chi_{c0} |  \psi^\dagger  (-\mbox{$\frac{i}{2}$} \tensor{\bf D} 
		\cdot \mbox{\boldmath $\sigma$}) \chi 
\chi^\dagger  (-\mbox{$\frac{i}{2}$} \tensor{\bf D} 
		\cdot \mbox{\boldmath $\sigma$}) \psi | \chi_{c0} \rangle \,.
\end{eqnarray}
\end{mathletters}

Using the relations (\ref{octS-id}) and (\ref{singP-scalar}),
the dominant terms in the decay rates for the 
$\chi_{cJ}$ reduce to
\begin{eqnarray}
\Gamma (\chi_{cJ})
& = &   {1 \over 2 M_{\chi_{cJ}}} \Bigg( 
{C^{(\underline{8},^3S)} \over m_c^2} \; 
\langle  \chi_{c0} | \psi^\dagger \mbox{\boldmath $\sigma$} T^a \chi \cdot 
	 \chi^\dagger \mbox{\boldmath $\sigma$} T^a \psi 
	 | \chi_{c0} \rangle^{(\mu)}
\nonumber \\
 & & \hspace{.5in}
\;+\; {C^{(\underline{1},^3P_J)}(\mu) \over m_c^4} \; 
\langle \chi_{c0} | \psi^\dagger (-\mbox{$\frac{i}{2}$} \tensor{\bf D} 
		\cdot \mbox{\boldmath $\sigma$}) \chi
	\chi^\dagger (-\mbox{$\frac{i}{2}$} \tensor{\bf D} 
		\cdot \mbox{\boldmath $\sigma$}) \psi
| \chi_{c0} \rangle \Bigg)  \;,
\label{Gam-chi}
\end{eqnarray}
where $C^{(\underline{1},^3P_J)}$ is a  linear combination
of the coefficients $C^{(\underline{1},^3P)}_n$, $n = 1,2,3$:
\begin{equation}
C^{(\underline{1},^3P_J)}(\mu)
\;=\; \delta_{J0} \; C^{(\underline{1},^3P)}_1
\;+\; {1 \over 3} \;  C^{(\underline{1},^3P)}_2(\mu)
\;+\; {(-1)^J  \over 3} \; C^{(\underline{1},^3P)}_3 \;.
\label{d-3P}
\end{equation}
In explicit form, these coefficients are
\begin{mathletters} \label{C-13P012}
\begin{eqnarray}
C^{(\underline{1},^3P_0)}(\mu)
&=&  {4 \pi \over 9} \alpha_s^2(2m_c) 
\left[ 1 
\;-\; {4 \over 27} \left( \ln{\mu \over 2 m_c} + {29 \over 6} \right) 
	n_f {\alpha_s \over \pi}\right]  \;,
\label{C-13P0}
\\
C^{(\underline{1},^3P_1)}(\mu)
&=&  {4 \pi \over 9} \alpha_s^2(2m_c) 
\left[ 0 \;-\; 
{4 \over 27} \left( \ln{\mu \over 2 m_c} + {4 \over 3} \right) 
	n_f {\alpha_s \over \pi}\right]  \;,
\label{C-13P1}
\\
C^{(\underline{1},^3P_2)}(\mu)
&=&  {4 \pi \over 9} \alpha_s^2(2m_c) 
\left[ {4 \over 15} \;-\; 
{4 \over 27} \left( \ln{\mu \over 2 m_c} + {29 \over 15} \right) 
	n_f {\alpha_s \over \pi}\right]  \;.
 \label{C-13P2}
\end{eqnarray}
\end{mathletters}
The constants $Q_J$ under the logarithm in these three expressions
are
\begin{equation}
Q_0 \;=\;  { 29 \over 6}   \;, \quad
Q_1 \;=\;  {  4 \over 3  } \;, \quad
Q_2 \;=\;  { 29 \over 15 } \;.
\label{QJ} 
\end{equation}

Up to corrections of relative order $v^2$,
the matrix elements in (\ref{Gam-chi}) are related to the standard
matrix elements defined in Ref. \cite{B-B-L} by a simple 
normalization factor:
\begin{eqnarray}
\langle \chi_{c0} | \psi^\dagger \mbox{\boldmath $\sigma$} T^a \chi \cdot
	\chi^\dagger \mbox{\boldmath $\sigma$} T^a \psi | \chi_{c0} \rangle
&\approx& 4 m_c 
    \;\langle \chi_{c0} |{\cal O}_8({}^3S_1)  | \chi_{c0} \rangle\,,
\\ 
\langle \chi_{c0} | \psi^\dagger (-\mbox{$\frac{i}{2}$} \tensor{\bf D} 
		\cdot \mbox{\boldmath $\sigma$}) \chi
	\chi^\dagger (-\mbox{$\frac{i}{2}$} \tensor{\bf D} 
		\cdot \mbox{\boldmath $\sigma$}) \psi | \chi_{c0} \rangle
&\approx& 12 m_c \; 
    \langle \chi_{c0} | {\cal O}_1({}^3P_{0})  | \chi_{c0} \rangle\,.
\end{eqnarray}
Using the vacuum-saturation approximation, the matrix element
$\langle \chi_{c0} | {\cal O}_1({}^3P_{0})  | \chi_{c0} \rangle$
can be expressed in terms of the derivative of the radial wavefunction 
for the $\chi_{c0}$ evaluated at the origin:
\begin{equation}
\langle \chi_{c0} | {\cal O}_1({}^3P_{0})  | \chi_{c0} \rangle
\; \approx \; 
{27 \over 2 \pi} |R_{\chi_{c0}}'(0)|^2 \;.
\end{equation}
The corrections are of relative order $v^4$.

We now compare our results for $C^{(\underline{1},^3P_J)}$
with previous calculations of these coefficients to order $\alpha_s^3$.
Our results agree with the $n_f \alpha_s^3$ terms obtained in 
recent calculations by Huang and Chao \cite{Huang-Chao} 
and by Petrelli \cite{Petrelli:PhD}, 
who also used dimensional regularization 
as the infrared cutoff for divergences associated with the emission 
of a soft gluon.  Our coefficients need not agree with those in
the original calculations by Barbieri et al. 
\cite{Barbieri:chi02,Barbieri:chi1}, since they used different 
infrared cutoffs.  However any differences in 
$C^{(\underline{1},^3P_J)}$ must be compensated by differences in 
the matrix element
$\langle \chi_{c0} | \psi^\dagger \mbox{\boldmath $\sigma$} T^a \chi \cdot 
	 \chi^\dagger \mbox{\boldmath $\sigma$} T^a \psi 
	 | \chi_{c0} \rangle$
in the factorization formula (\ref{Gam-chi}).
This implies that the differences between the coefficients
$C^{(\underline{1},^3P_J)}$ obtained with two different infrared cutoffs 
must be independent of $J$.

In Ref. \cite{Barbieri:chi02},
Barbieri et al. calculated the coefficients of the color-singlet 
matrix element in the decay rates of the $\chi_{c0}$ and $\chi_{c2}$ 
using the binding energy of the $c \bar c$ pair to
regularize the infrared divergences that arise from the emission 
of a soft gluon.  Taking the annihilating $c \bar c$ pair to have 
invariant mass $M$, the infrared divergence shows up as a logarithm of 
$M^2 - 4 m_c^2$.  This cutoff can be translated into an equivalent
cutoff $\Lambda$ on the momentum of the soft gluon by replacing
$\ln(4m_c^2/(M^2 - 4 m_c^2)) \to - \ln (\Lambda/m_c)$. 
In Ref. \cite{Barbieri:chi1},
Barbieri et al. calculated the decay rates of 
$\chi_{c0}$, $\chi_{c1}$, and $\chi_{c2}$ into two-jet configurations
defined by an angular resolution $\delta$ and a fractional energy resolution
$\epsilon$.  Their result for $\chi_{c1}$ is independent of $\delta$ 
and depends logarithmically on $\epsilon$. 
This cutoff can be translated into an equivalent infrared
cutoff $\Lambda$ on the momentum of the soft gluon by replacing
$\ln(2 \epsilon) \to \ln (\Lambda/m_c)$.
Expressing the results of Barbieri et al. in the same form as in 
(\ref{C-13P012}), the expressions inside the parentheses
are $ \ln(\Lambda/m_c) + Q_J'$, where the numbers $Q_J'$ are 
\begin{equation}
Q_0' \;=\;  4   \;, \quad
Q_1' \;=\;  {1 \over 2} \;, \quad
Q_2' \;=\;  {11 \over 10} \;.
\label{QJ:Barbieri} 
\end{equation}
Comparing the coefficients in (\ref{QJ}) and (\ref{QJ:Barbieri}),
we see that $Q_J - Q_J' = {5 \over 6}$, independent of $J$.  Thus the 
differences between the two calculations correspond simply to 
different choices for the definition of the color-octet matrix element.
The relation between the matrix elements in the two calculations is 
\begin{eqnarray}
\langle \chi_{c0} |  \psi^\dagger \mbox{\boldmath $\sigma$} T^a \chi \cdot
	\chi^\dagger \mbox{\boldmath $\sigma$} T^a \psi 
	| \chi_{c0} \rangle^{(\mu)} \Bigg|_{\rm dim.reg.}
&\; = \;&
\langle \chi_{c0} | \psi^\dagger \mbox{\boldmath $\sigma$} T^a \chi \cdot
	\chi^\dagger \mbox{\boldmath $\sigma$} T^a \psi 
	| \chi_{c0} \rangle^{(\Lambda)} \Bigg|_{\rm cutoff}
\nonumber \\
&& \hspace{-1.5in}
\;+\; {16 \alpha_s \over 81 \pi m_c^2}
\left( \ln {\mu \over 2 \Lambda}  + {5 \over 6} \right) 
\langle \chi_{c0} | \psi^\dagger (-\mbox{$\frac{i}{2}$} \tensor{\bf D} 
		\cdot \mbox{\boldmath $\sigma$}) \chi
	\chi^\dagger (-\mbox{$\frac{i}{2}$} \tensor{\bf D} 
		\cdot \mbox{\boldmath $\sigma$}) \psi | \chi_{c0} \rangle  \,.
\label{matel:diff}
\end{eqnarray}
This identification can be verified by repeating the calculation 
of the coefficient $C^{(\underline{1},^3P)}_2$ in Section III
using an infrared momentum cutoff $\Lambda$.  
On the QCD side of the matching condition,
one must put a cutoff $x > \Lambda/m_c$ on the integral in 
(\ref{int-x}) for $n=0$.  On the NRQCD side, one must put a 
cutoff $k > \Lambda$ on the integral in (\ref{int-k}).
The relation (\ref{matel:diff}) 
is identical to the  relation between the corresponding 
production matrix elements defined by dimensional regularization
and by a momentum cutoff, which  was found in Ref. \cite{Braaten-Chen:dimreg}.

In Ref. \cite{Mangano-Petrelli} and \cite{Petrelli}, Mangano and
Petrelli used the next-to-leading order results of Barbieri et al. 
for $\chi_{c0}$ and $\chi_{c2}$, but they performed an independent
calculation for $\chi_{c1}$.  Their result corresponds to 
$Q_1' = -{7 \over 3}$.  
Since this result disagrees with four other independent calculations,
we conclude that it is incorrect.

\section{Conclusions}

In the NRQCD factorization framework, the short-distance coefficients
in annihilation decay rates and in inclusive production cross sections
can be calculated systematically as a power series in $\alpha_s(m_c)$.
For most applications, next-to-leading order calculations are essential
for accurate predictions.
Dimensional regularization is the most convenient method for regularizing 
the infrared and ultraviolet divergences that arise in calculations
beyond leading order in $\alpha_s$. 
The  generalization of the threshold expansion method to $N$ 
dimensions allows dimensional regularization to be used consistently in 
quarkonium calculations. In this paper, we used this method 
to calculate the terms proportional to $n_f \alpha_s^3$
in the short-distance coefficients of color-singlet matrix elements
in the annihilation decay rates of P-wave charmonium states, 
thus resolving the discrepancies between previous calculations.  
Our results agree with the original 
calculations by Barbieri et al. after allowing for differences in the 
color-octet matrix element due to different infrared cutoffs.
Our results also agree with recent calculations using covariant 
projection methods in conjunction with dimensional regularization.
In conclusion, the threshold expansion method combined with dimensional 
regularization provides a general and powerful method for 
carrying out quarkonium calculations beyond leading order in $\alpha_s$.

\acknowledgements

This work was supported in part by the U.S.
Department of Energy, Division of High Energy Physics, under 
Grant DE-FG02-91-ER40684.

\appendix

\section{Nonrelativistic expansion of spinors}

In this Appendix,  we present the formulas for 
the nonrelativistic approximations of spinors that are needed to 
calculate short-distance coefficients for  heavy 
quarkonium decay rates using the threshold expansion method. 
All the formulas below hold in $N$ spatial dimensions. 
The representation for gamma matrices that is most convenient 
for carrying out the nonrelativistic expansion of a spinor 
is the Dirac representation:
\begin{equation}
\gamma^0  \;=\; 
\left( \begin{array}{cc} 
	1 &  0  \\ 
	0 & -1 
	\end{array} \right) \;, 
\qquad
\gamma^i \;=\;  
\left( \begin{array}{cc} 
	     0    & \sigma^i \\ 
	-\sigma^i &    0        
	\end{array} \right) \;.
\end{equation}
In the  center-of-momentum (CM) frame of the $c\bar{c}$ pair,  
their momenta $p$ and $\bar p$ of the $c$ and $\bar c$ can be written
\begin{mathletters}
\begin{eqnarray}
p &\;=\;& ( E_q,   {\bf q} ) \;,
\\
\bar p &\;=\;& ( E_q,   - {\bf q} )  \;,
\end{eqnarray}
\end{mathletters}
where $ E_q = \sqrt{m_c^2 + {\bf q}^2 } $. 
The spinors for the $c$ and 
the $\bar c$ in the CM frame are
\begin{mathletters}
\label{spin-rest}
\begin{eqnarray}
u(p) 
&\;=\;& {1 \over \sqrt{E_q + m_c}}
\left( \begin{array}{c} 
	(E_q + m_c) \; \xi \\ 
	{\bf q} \cdot \mbox{\boldmath $\sigma$} \; \xi 
	\end{array} \right) \;,
\label{u-rest}
\\
v(\bar p) 
&\;=\;& {1 \over \sqrt{E_q + m_c}}
\left( \begin{array}{c} 
	- {\bf q} \cdot \mbox{\boldmath $\sigma$} \; \eta \\
	(E_q + m_c) \; \eta
	\end{array} \right) \;.
\label{v-rest}
\end{eqnarray}
\end{mathletters}
Color and spin quantum numbers on the Dirac spinors
and on the Pauli spinors $\eta$ and $\xi$ are suppressed.
If the 
Pauli spinors are normalized so that $\eta^\dagger \eta = \xi^\dagger \xi = 1$,
the spinors in (\ref{spin-rest}) satisfy
 $\bar u u = - \bar v v = 2 m_c$. 

The independent quantities that can be formed by sandwiching
3 or fewer Dirac matrices between $\bar v(\bar p) $ and $u(p)$ are
\begin{mathletters}
\label{bispinors}
\begin{eqnarray}
&& \bar v(\bar p) u(p) \;=\; 
- 2 \; \eta^\dagger ({\bf q} \cdot \mbox{\boldmath $\sigma$}) \xi ,
\\
&& \bar v(\bar p) \gamma^\mu u(p) \;=\; - g^{\mu j} \;
\left( 2 E_q \; \eta^\dagger \sigma^j \xi \;-\; {2 \over E_q + m_c} \; q^j \;
	\eta^\dagger ( {\bf q} \cdot \mbox{\boldmath $\sigma$} ) \xi \right) 
\;,                                                                              
\label{bispinor-gam}
\\ 
&& \bar v(\bar p) ( \gamma^\mu \gamma^\nu
           -\gamma^\nu \gamma^\mu ) u(p) \;=\; 
  4 ( g^{\mu 0} g^{\nu j} - g^{\nu 0} g^{\mu j} ) 
\left( { m_c } \; \eta^\dagger \sigma^j \xi 
	\;+\; {1 \over E_q + m_c } q^j \; 
	\eta^\dagger ({\bf q} \cdot \mbox{\boldmath $\sigma$}) \xi \right)
\nonumber \\
&& \hspace{2in}
\;+\;  g^{\mu j} \; g^{\nu k} \; 
	\eta^\dagger \{ [ \sigma^j, \sigma^k ], 
		{\bf q} \cdot \mbox{\boldmath $\sigma$} \} \xi \;,
\\
&& \bar v(\bar p) ( \gamma^\mu \gamma^\nu \gamma^\mu
                   -\gamma^\mu \gamma^\nu \gamma^\mu ) u(p) 
\nonumber \\
&& \hspace{.5in} 
\;=\;  g^{\mu i} \; g^{\nu j} \; g^{\mu k} 
\Bigg(  E_q \; \eta^\dagger \{ [ \sigma^i, \sigma^j ] , \sigma^k \} \xi 
\;-\; {q^i \over E_q + m_c} \; \eta^\dagger \{ [ \sigma^j, \sigma^k ] , 
			{\bf q} \cdot \mbox{\boldmath $\sigma$} \} \xi 
\nonumber \\ 
&& \hspace{1.5in}
\;-\; {q^j \over E_q + m_c} \; \eta^\dagger \{ [ \sigma^k, \sigma^i ] , 
			{\bf q} \cdot \mbox{\boldmath $\sigma$} \} \xi 
\;-\; {q^k \over E_q + m_c} \; \eta^\dagger \{ [ \sigma^i, \sigma^j ] , 
			{\bf q} \cdot \mbox{\boldmath $\sigma$} \} \xi 
\Bigg)
\nonumber \\
&& \hspace{1in}
\;+\; {4} 
\left(     g^{\mu 0} g^{\nu i} g^{\mu j} 
	+  g^{\nu 0} g^{\mu i} g^{\mu j}
	+  g^{\mu 0} g^{\mu i} g^{\nu j}  
 \right)
\left( \eta^\dagger q^i \sigma^j \xi 
	\;-\; \eta^\dagger q^j \sigma^i \xi \right) \;.
\end{eqnarray}
\end{mathletters}
These expressions can be obtained from the corresponding spinor
factors in Appendix A of Ref.\cite{Braaten-Chen:dimreg} by setting 
$L^\mu_{\ j} =- g^{\mu j } = \delta ^{\mu j } $ 
and $ P^ \mu = 2 E_q g^{\mu 0 } $ and by taking the hermitian conjugate. 
Using the expressions for the spinor factors in (\ref{bispinors}), 
it is easy to carry out their nonrelativistic 
expansions in powers of ${\bf q}$.


\vfill \eject

\begin{figure}
{ Fig.~1. 
The lowest-order cut Feynman diagram in QCD that are associated with  the 
process $c \bar c \to q \bar q$. }
\end{figure}

\begin{figure}
{ Fig.~2. 
The lowest-order cut Feynman diagrams in QCD that are associated with  the 
process $c \bar c \to q \bar q g$. }
\end{figure}

\begin{figure}
{ Fig.~3. 
The lowest-order cut Feynman diagrams in QCD that are associated with the 
process $c \bar c \to g g$. }
\end{figure}

\begin{figure}
{ Fig.~4. 
The lowest-order Feynman diagram for NRQCD matrix elements such as 
$\langle c \bar c' | \psi^\dagger \mbox{\boldmath $\sigma$} T^a \chi \cdot
	\chi^\dagger \mbox{\boldmath $\sigma$} T^a \psi | c \bar c \rangle$.}
\end{figure}

\begin{figure}
{Fig.~5.
The next-to-leading-order Feynman diagrams for the NRQCD matrix element
$\langle c\bar{c}' | \psi^\dagger \mbox{\boldmath $\sigma$} T^a \chi \cdot
	\chi^\dagger \mbox{\boldmath $\sigma$} T^a \psi | c\bar{c} \rangle$
when the initial and final $c \bar c$ pairs are  in color-singlet states.}
\end{figure}

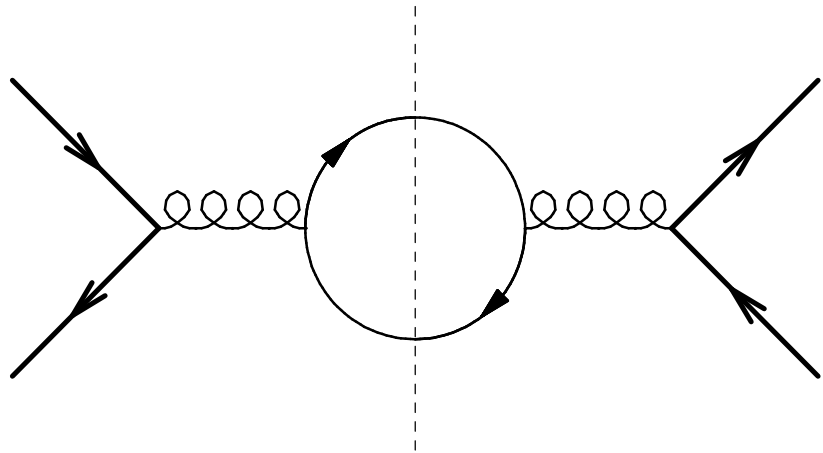
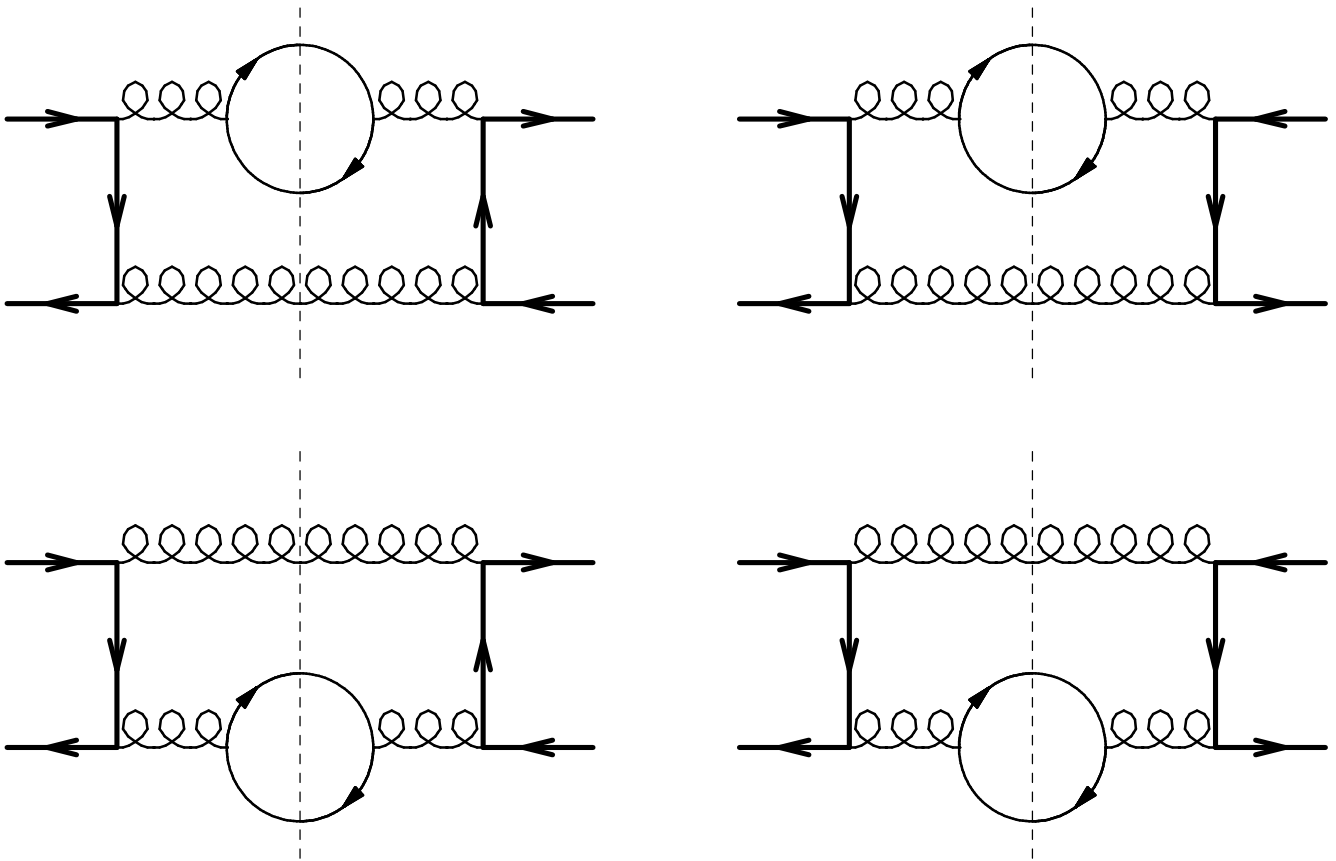
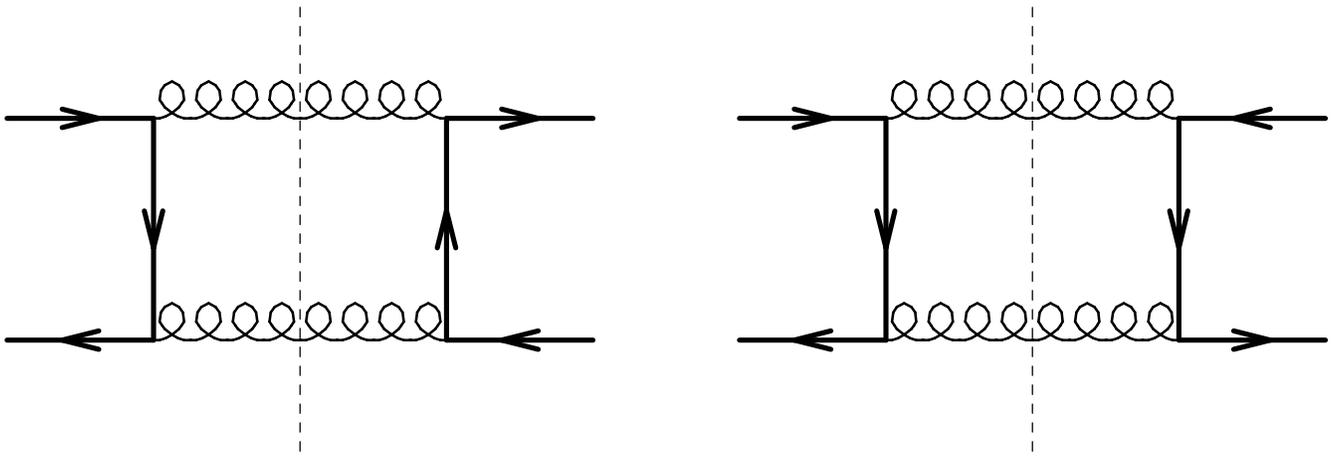
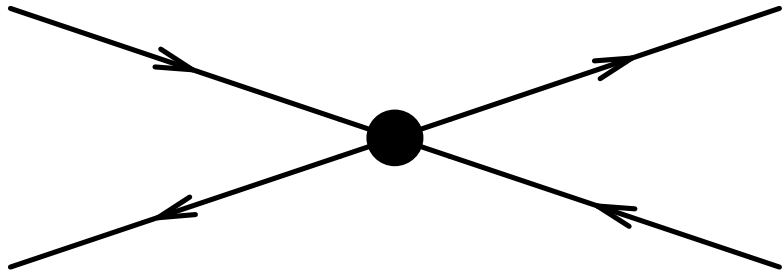
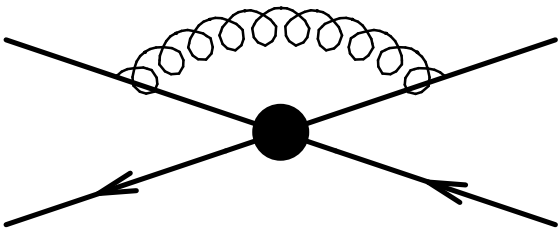
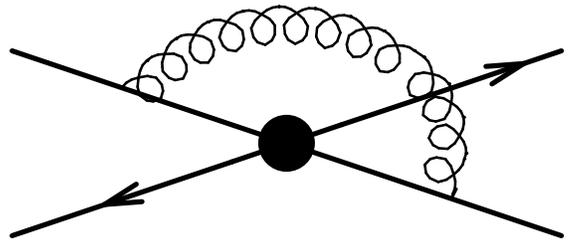
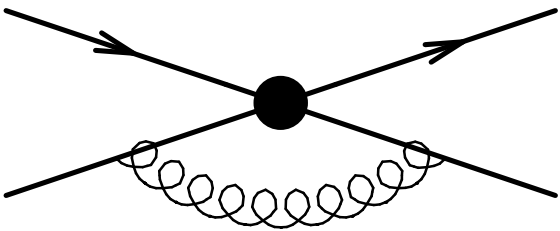
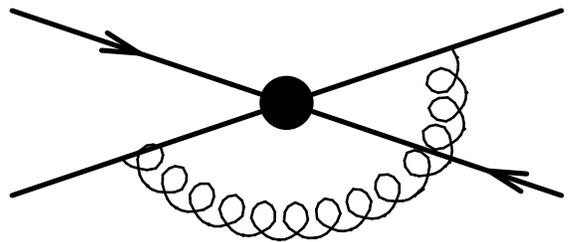
\end{document}